\documentclass[letterpaper,twocolumn,10pt]{article}
\usepackage{usenix2019_v3}

\usepackage{lineno}
\usepackage{amsmath,amsfonts,amssymb}
\usepackage{graphicx}
\usepackage{cite}
\usepackage{tcolorbox}
\usepackage{algorithm}
\usepackage{listings}
\usepackage{booktabs}
\usepackage{tikz}
\usepackage{tablefootnote}
\usepackage[flushleft]{threeparttable}

\usepackage{url}

\usepackage{breakurl}
\usepackage{hyperref}

\lstdefinestyle{asmstyle}{
  belowcaptionskip=1\baselineskip,
  breaklines=true,
  frame=None,
  xleftmargin=\parindent,
  language=[x86masm]Assembler,
  showstringspaces=false,
  basicstyle=\footnotesize\ttfamily,
  commentstyle=\itshape\color{purple!40!black},
  keywordstyle=\bfseries\color{green!40!black},
  identifierstyle=\color{blue},
  numbers=left,
  numberstyle=\tiny\color{gray},
}

\definecolor{mGreen}{rgb}{0,0.6,0}
\definecolor{mGray}{rgb}{0.5,0.5,0.5}
\definecolor{mPurple}{rgb}{0.58,0,0.82}
\definecolor{backgroundColour}{rgb}{0.95,0.95,0.92}
\usepackage[noend]{algpseudocode}
\usepackage{filecontents}

\lstdefinestyle{CStyle}{
    backgroundcolor=\color{backgroundColour},   
    commentstyle=\color{mGreen},
    keywordstyle=\color{magenta},
    numberstyle=\tiny\color{mGray},
    stringstyle=\color{mPurple},
    basicstyle=\footnotesize,
    breakatwhitespace=false,         
    breaklines=true,                 
    captionpos=b,                    
    keepspaces=true,                 
    numbers=left,                    
    numbersep=5pt,                  
    showspaces=false,                
    showstringspaces=false,
    showtabs=false,                  
    tabsize=2,
    language=C
}

\definecolor{lightgray}{HTML}{F7F7F7}
\definecolor{customblack}{HTML}{A0A0A0}

\newcommand*\circled[1]{\tikz[baseline=(char.base)]{
            \node[shape=circle,draw,inner sep=1pt] (char) {#1};}}

\newcommand*\circleB[1]{\tikz[baseline=(char.base)]{\node[shape=circle,fill,inner sep=1pt] (char) {\textcolor{white}{#1}};}}

\begin{document}

\date{}

\title{\Large \bf Shesha\thanks{Shesha or Sheshanaga is a serpentine demigod in Hindu mythology with multiple heads and serves as the celestial bed of Lord Vishnu.}~~: Multi-head Microarchitectural Leakage Discovery \\ in new-generation Intel Processors}


\author{
{\rm Anirban Chakraborty}\\
Indian Institute of Technology Kharagpur    \\
\href{mailto:ch.anirban00727@gmail.com}{anirban.chakraborty@iitkgp.ac.in}
\and
{\rm Nimish Mishra}\\
Indian Institute of Technology Kharagpur    \\
\href{mailto:neelam.nimish@gmail.com}{nimish.mishra@kgpian.iitkgp.ac.in}
\and
{\rm Debdeep Mukhopadhyay}\\
Indian Institute of Technology Kharagpur    \\
\href{mailto:debdeep.mukhopadhyay@gmail.com}{debdeep@cse.iitkgp.ac.in}
}


\maketitle

\begin{abstract}
Transient execution attacks have been one of the widely explored microarchitectural side channels since the discovery of Spectre and Meltdown. However, much of the research has been driven by manual discovery of new transient paths through well-known speculative events. Although a few attempts exist in literature on automating transient leakage discovery, such tools focus on finding variants of known transient attacks and explore a small subset of instruction set. Further, they take a random fuzzing approach that does not scale as the complexity of search space increases. In this work, we identify that the search space of bad speculation is disjointedly fragmented into \textit{equivalence classes}, and then use this observation to develop a framework named {\sf Shesha}, inspired by Particle Swarm Optimization, which exhibits faster convergence rates than state-of-the-art fuzzing techniques for automatic discovery of transient execution attacks. We then use {\sf Shesha} to explore the vast search space of extensions to the \texttt{x86} Instruction Set Architecture (ISAs), thereby focusing on previously unexplored avenues of bad speculation. As such, we report five previously unreported transient execution paths in Instruction Set Extensions (ISEs) on new generation of Intel processors. We then perform extensive reverse engineering of each of the transient execution paths and provide root-cause analysis.
Using the discovered transient execution paths, we develop attack building blocks to exhibit exploitable transient windows. Finally, we demonstrate data leakage from Fused Multiply-Add instructions through SIMD buffer and extract victim data from various cryptographic implementations.

\end{abstract}

\section{Introduction}  \label{sec:intro}
With a plethora of sophisticated optimizations incorporated in modern processors, speculative execution plays a crucial role in determining the overall performance of the system. However, in $2018$ a new class of attacks, called \emph{transient attacks} introduced with Spectre~\cite{kocher2020spectre} and Meltdown~\cite{lipp2020meltdown}, have brought into perspective the side-channel implication of speculative execution. These attacks rely on software-accessible side-channels to extract secrets by forcing the CPU to enter into \emph{transient state} and leave microarchitectural traces. A number of attacks have been demonstrated in literature that exploit these vulnerabilities to leak secret information across security boundaries~\cite{barberis2022branch, canella2019fallout, wikner2022retbleed, ragab2021crosstalk, schwarz2019zombieload, stecklina2018lazyfp, van2018foreshadow, van2020lvi, van2019ridl, moghimi2023downfall, bhattacharyya2019smotherspectre, canella2019systematic, chen2019sgxpectre, goktas2020speculative, kiriansky2018speculative, koruyeh2018spectre, maisuradze2018ret2spec, schwarz2019netspectre, trippel2018meltdownprime, van2020sgaxe, van2021cacheout, weisse2018foreshadow, xiong2020survey, chakraborty2022timed}. 
Intel refers to such cases as \emph{bad speculation}~\cite{intel_sdm} where the CPU is being coerced into a condition where the issued micro-operations ($\mu$Ops) have to be discarded while their results are transiently reflected in microarchitectural states.

As per Intel, bad speculation happens due to three broader causes - branch misprediction, microcode assists and machine clear~\cite{intel_sdm}. While a number of works have explored branch misprediction-based transient attacks in literature~\cite{canella2019systematic, chen2019sgxpectre, kocher2020spectre, koruyeh2018spectre, maisuradze2018ret2spec}, very little research has been made to explore machine clears and microcode assists.
Recently, authors in \cite{ragab2021rage} explored the occurrences of machine clears in modern processors and demonstrated multiple transient execution paths based on different machine clear events. The findings of \cite{ragab2021rage} open up a new frontier for transient execution discoveries that motivate further exploration. While most of the known works on transient leakage discoveries have been the results of manual effort, recent literature has seen quite some attempts to automate such leakage exploration~\cite{moghimi2020medusa, hofmann2023speculation, weber2021osiris, ibrahim2022microarchitectural, xiao2019speechminer}.  However, these tools were developed to target certain classes of speculations and cannot be directly extended to explore the vast space of bad speculation (refer Section~\ref{sec:related_works} for a detailed discussion and comparison with our tool). 
Additionally, due to the large number of x86 instructions, the prior works on automation only consider a small subset of Instruction Set Extensions (ISE), thereby missing out on a large pool of specialized instructions that are performance-bound and share various hardware resources. In fact, some of these ISEs, especially those concerned with SIMD, streaming SIMD, and Vector extensions, have their own separate data path (i.e. execution units), registers, temporal buffers, etc, making them an ideal spot for enabling speculation, and possibly bad speculation. Evidently, a systematic exploration of bad speculation in the context of ISEs is necessary to understand transient execution related vulnerabilities of ISEs. Concretely, in this work, we deal with the following question:

\smallskip

\textit{Given the wide-array of ISEs, can we develop a generic, better-than-random (i.e. better than fuzzing) approach to automatically explore the vast search space of bad speculation across ISE executions and develop previously unexplored transient executions paths?}

\smallskip

In this work, we choose to adopt an alternative path than random fuzzing for the following reasons:
\begin{enumerate} \itemsep0em
    \item \textbf{Undirected vs Directed searches}: Fuzzing without feedback (like coverage) is an undirected approach that cannot incrementally develop upon good solutions since it applies random mutations without any sense of direction. In contrast, directed approaches (like evolutionary algorithms) consistently try to make good solutions \textit{better}. This ensures faster convergence for directed approaches as opposed to undirected fuzzing-based approaches.

    \item \textbf{Scalability with vastly expressive search spaces}: Randomized, undirected fuzzing leads to scalability issues with vastly expressive search spaces. With modern ISAs (along with extensions), the total number of possible instructions exceeds $16000$. Random fuzzing through such expansive search spaces causes problems, as evident in prior works in the literature. For instance, works like \cite{hofmann2023speculation, oleksenko2023hide, moghimi2020medusa} limit the extent of their instruction set to prevent blowing up the search space.
\end{enumerate}

Given these observations, we choose to develop a guided approach that will not require restricting the search space, thereby increasing chances of finding corner cases of bad speculation. We develop a framework- \textsf{Shesha} - based on the principles of \emph{Particle Swarm Optimizers (PSO)}~\cite{kennedy1995particle} to allow for a directed exploration of the ISE search space, and faster convergence on code sequences enabling bad speculation. We choose PSO as the foundation of \textsf{Shesha} since PSOs allow for a delicate balance between choosing from local and global gains in the search space. However, unlike a textbook PSO, we observe that the search space for bad speculation is disjointedly fragmented, and therefore allows modifications to the generic PSO that enable faster convergence. Consequently, \textsf{Shesha} is able to uncover newer, previously unreported avenues of bad speculation in ISEs: \circled{1} SIMD-Vector instruction intermixing (which has been unobserved~\cite{ragab2021rage} in newer generation Intel processors prior to Alder Lake and Sapphire Rapids), \circled{2} execution of fused-multiply-add (FMA) instructions, \circled{3} intermixing single and double precision in ISEs, and \circled{4} performing denormal arithmetic on combination of AES-NI and SIMD instruction sets. We perform extensive reverse engineering of each of these avenues to uncover the cause of bad speculation in each of these cases, which helps us to construct abstract code sequence descriptions called \textit{gadgets}, vulnerable to transient execution attacks. 
To summarize, we make the following contributions:
\begin{itemize}     \itemsep0em
    \item We propose \textsf{Shesha}, a framework to explore previously untapped search space of bad speculation with respect to \textit{Instruction Set Extensions}. Our methodology relies on a modified version of Particle Swarm Optimization and exploits the disjointedly fragmented nature of transient execution attacks to deliver faster convergence rates.

    \item We uncover newer and previously unreported avenues of transient execution paths: SIMD-Vector transitions, execution of fused-multiply-add (FMA) instructions, intermixing single and double precision in ISEs, and performing denormal arithmetic on a combination of AES-NI and SIMD instruction sets. We also provide extensive root-cause analysis for each of the scenarios.

    \item We characterize and evaluate the leakage strengths of each of the uncovered transient execution paths and discover SIMD-Vector transitions to exhibit the \textit{longest} adversarial-friendly window of execution amongst all transient execution paths. We also demonstrate Simultaneous Multi-Threaded covert channels as well as Load Value Injection (LVI) attacks on vulnerable targets.

    \item Finally, we reverse-engineer the Fused Multiply-Add (FMA) execution unit to root-cause the transient leakage and exploit it to leak victim data from multiple cryptographic implementations. Our findings suggest that data used by \emph{all} FMA and IFMA instructions are prone to leak across hyperthreads, rendering the applications using FMA for optimization vulnerable.
\end{itemize}

The rest of the paper is organized as follows: Section~\ref{sec:background} provides the necessary background information. Section~\ref{sec:automatic_analysis}, \ref{sec:transient_execution_paths} and \ref{sec:ablation_study} are dedicated to the \textsf{Shesha} framework and its discoveries. In Section~\ref{sec:root_cause}, we perform a root-cause analysis of the discovered transient execution paths and then discuss how they can aid as building blocks to micro-architectural attacks in Section~\ref{sec:attack_building_blocks}. In Section~\ref{sec:fma_leaking}, we show practical exploitation of leakage from FMA instructions. We then provide a discussion on related works in Section~\ref{sec:related_works} and mitigation strategies in Section~\ref{sec:mitigations}. Finally, we conclude in Section~\ref{sec:conclusion}.

\smallskip
\noindent \textbf{Responsible Disclosure:} We have disclosed the results of this work, including the leakage due to FMA execution unit to Intel. Intel acknowledged the leakage from the FMA execution unit, and responded that mitigation from~\cite{intel_gather} could cover this issue as well; therefore, no embargo period is required.

\smallskip
\noindent \textbf{Artifact Availability:} The tool and proof-of-concept attack codes can be found at \url{https://github.com/SEAL-IIT-KGP/shesha}.

\color{black}
\section{Background}
\label{sec:background}

\subsection{Superscalar Computing}
High-performance CPUs, such as Intel Core and Xeon processors, utilize multiple cores, high-speed memory units, and support parallel processing with efficient isolation and security measures. 
Memory isolation is achieved through individual per-process virtual address spaces. These virtual addresses are translated to page table entries (PTEs) containing data location, access control, and status bits for access control. 
These modern CPUs employ Simultaneous Multithreading (SMT) that allows multiple threads to execute on the same core, providing architectural isolation. Intel CPUs support two threads to run simultaneously per physical core. Another important feature called speculative execution enables the CPU core to execute instructions in the pipeline even when there are dependencies on prior unresolved operations. In case of incorrect predictions, the CPU corrects itself by re-executing instructions. While architecturally invisible, there may be observable side effects due to microarchitectural state changes. In this context, such an instruction is referred to as a \emph{transient instruction}~\cite{canella2019systematic,kocher2020spectre,lipp2020meltdown}. Transient execution attacks capitalize on these transient instructions to leak sensitive data across security boundaries.
Single Instruction, Multiple Data (SIMD) enables data-level parallelism, performing the same operation on multiple sets of data. AVX2 and AVX-512 are key SIMD extensions for x86 architecture. Finally, the CPU includes a shared last-level cache (LLC) across execution cores and an interconnect bus linking the LLC, cores, and DRAM. Additionally, each core has its own private caches - L1 and L2 caches. Temporal buffers are employed to optimize micro-operations, such as the fill buffer for fetching cache line data bits and the store buffer for holding data before committing it to the cache.


\subsection{Particle Swarm Optimization}
\label{sec:particle_swarm_optimization}
A Particle Swarm Optimizer~\cite{kennedy1995particle}, abbreviated PSO, is a generic evolutionary method suited to solving optimization problems. Unlike Genetic Algorithms that rely upon evolution of genetic information, PSOs work on the delicate interplay of individual and swarm behaviours in nature (like ant swarm, bee swarm etc). Abstractly, a PSO balances two opposing forces- \textit{exploration} and \textit{exploitation} in order to optimize an objective function. Here, \textit{exploration} abstracts the random choices that encourage the PSO for a random walk across the search space in hopes of finding newer paths to global optimums. On the other hand, \textit{exploitation} abstracts the greedy choices, encouraging the PSO to focus on a direct path to currently known optimums. As evident, too much of \textit{exploration} is no better than a random search, while too much of \textit{exploitation} risks getting stuck in local optimums. A PSO balances the two forces by switching back-and-forth between \textit{particle} and \textit{swarm} behaviour. A particle performs random searches across the search space, and if optimal paths are discovered, broadcasts that information to other particles, such that the entire swarm (where swarm can be interpreted as the set of all particles) starts to use the information from that one particle. Consequently, the particle behaviour abstracts exploration, while the swarm behaviour abstracts exploitation.

Syntactically, a PSO is characterized by a \textit{fitness} function $f: \mathcal{R}^{n} \rightarrow \mathcal{R}$, mapping an $n$-dimensional vector (termed \textit{particle position}, or $\mathbf{x}$) to a real-value (termed \textit{fitness}) which expresses how close the particle position is wrt. global optimum. Initially, the PSO samples a random \textit{population} of $N$ particles, denoted by a set of position vectors $\mathbb{X} = \{\mathbf{x}_{1}, \mathbf{x}_{2}, \mathbf{x}_{3}, ..., \mathbf{x}_{N}\}$. For an iteration $i$ of the PSO algorithm and particle $j: \{1 \le j \le N\}$, the terms \emph{velocity} and \emph{position} are defined as:
\begin{equation*}
\begin{split}
    \mathbf{v}_{j}(i+1) &= \alpha . \mathbf{v}_{j}(i) + \beta . (\mathbf{p}_{j}(i) - \mathbf{x}_{j}(i)) + \gamma . (\mathbf{p}_{g}(i) - \mathbf{x}_{j}(i)) \\
    \mathbf{x}_{j}(i+1) &= \mathbf{x}_{j}(i) + \mathbf{v}_{j}(i+1)
    \label{position_equation}
\end{split}
\end{equation*}
where $\alpha$ is the inertial weight, $\beta$ is the cognitive acceleration coefficient, $\gamma$ is the social acceleration coefficient, $\mathbf{p}_{j}$ is the best position (where the fitness function is maximally optimal, up to iteration $i$) of particle $j$, and $\mathbf{p}_{g}$ is the best position for the entire swarm. Abstractly, in every iteration $i+1$, each particle $j$ attempts to \textit{update} its position $\mathbf{x}_{j}(i+1)$ based on the position in the previous iteration $\mathbf{x}_{j}(i)$ and a computed \textit{velocity} $\mathbf{v}_{j}(i+1)$. Now, this velocity vector is computed by combining three pieces of information:
\begin{enumerate}   \itemsep0em
    \item \textbf{Velocity in the previous iteration}: denoted by $\mathbf{v}_{j}(i)$
    \item \textbf{Exploration / Particle behaviour}: Velocity computed by considering the best position explored by the particle so far (denoted by the vector $\mathbf{p}_{j}(i)$).
    \item \textbf{Exploitation / Swarm behaviour}: Velocity computed by considering the best position across the entire swarm (denoted by the vector $\mathbf{p}_{g}(i)$).
\end{enumerate}
The parameters $\alpha$, $\beta$, and $\gamma \in [0, 1]$ control the importance given to each of these three information points, and directly affect how a particle moves across the search space.

\begin{table*}[!t]
\centering
\caption{Performance counters reporting on occurrences of bad speculation. (TMA: Top-down Microarchitecture Analysis)}
\label{tab:pmc}
\resizebox{0.97\textwidth}{!}{%
\begin{tabular}{|c|c|c|}
\hline
\textbf{Name} &  \textbf{Description} & \textbf{Class of bad speculation}\\ \hline
ASSISTS.FP & Counts all microcode Floating Point assists. & Microcode assist \\ \hline
ASSISTS.HARDWARE & Counts all hardware assists. & Microcode assist \\ \hline
ASSISTS.PAGE\_FAULT & Counts all assists related to page faults. & Microcode assist \\ \hline
ASSISTS.SSE\_AVX\_MIX & Counts all assists related to AVX-SSE transitions. & Microcode assist \\ \hline
MACHINE\_CLEARS.DISAMBIGUATION & Counts machine clears due to disambiguation issues & Machine clear \\ \hline
MACHINE\_CLEARS.MEMORY\_ORDERING & Counts machine clears due to memory ordering issues & Machine clear \\ \hline
MACHINE\_CLEARS.SMC & Counts machine clears due to self-modifying code & Machine clear \\ \hline
BR\_MISP\_RETIRED.ALL\_BRANCHES & Counts retired, mispredicted branch instructions. & Branch misprediction \\ \hline
TOPDOWN.BR\_MISPREDICT\_SLOTS & Counts TMA slots wasted due to branch misprediction & Branch misprediction \\ \hline
\end{tabular}%
}
\end{table*}

\section{\textsf{Shesha}: Automatic Analysis of Bad Speculation in Instruction Set Extensions}
\label{sec:automatic_analysis}
In this section, we put forward a Particle Swarm Optimizer (PSO) inspired framework to perform an optimized exploration of search space related to bad speculation in ISEs. Towards this, we first elaborate on how the search space of bad speculation is \emph{disjointedly fragmented}, allowing us to make modifications to generic PSO semantics (cf. Section~\ref{sec:particle_swarm_optimization}) that provide faster convergence rates. Concretely, following the semantics of Section~\ref{sec:particle_swarm_optimization}, instead of relying upon the entire swarm for swarm behaviour, the algorithm need only rely upon fragmented portions of the swarm, thereby allowing more granularity and enabling faster convergence. The disjointedly fragmented nature of the search space allows splitting \textsf{Shesha} into two phases: \circled{1} Cognitive phase (where there is no social acceleration), and \circled{2} Mixed phase (partial social/cognitive acceleration). By end of phase \circled{1}, each particle of the swarm is present in one of the disjointedly fragmented sub-space, and has knowledge of other particles in the same subspace. Thereby, in phase \circled{2}, the social acceleration component can be focused only on the fragmented sub-space. We detail this modification to the PSO semantics further in this section, along with the threat model.

\subsection{Threat Model}
Since the objective of \textsf{Shesha} is to \textit{explore} possible avenues of transient execution attacks, we assume Ring 3 \texttt{sudo} privileges. However, for attack demonstration, we do not require \texttt{sudo} privilege and only assume user-level privilege. We further assume x86-64-based target systems with latest microcode patches and secure (patched) operating system, with all state-of-the-art mitigations against transient execution attacks. We assume no publicly known exploitable vulnerability is present in the system. We specifically focus on Intel CPUs with newer generation microarchitecture, both on client and server platforms.
For our experiments, we execute \textsf{Shesha} on Intel CPUs from four generations: Alder Lake(12th gen Core i7) and Comet Lake (11th gen Core i7) on the client side and Sapphire Rapids (4th gen Xeon) and Ice Lake (3rd gen Xeon) on the server side. All subsequent discussions are contextual to these aforementioned machines.

\subsection{Fitness function}
\label{sec:fitness_function}
As discussed in Section~\ref{sec:particle_swarm_optimization}, the objective of a PSO is to optimize a well-defined fitness function $f: \mathcal{R}^{n} \rightarrow \mathcal{R}$. Since our search space is instruction sequences enabling bad speculation over ISEs, our fitness function must be a combination of indicators that capture the existence or absence of bad speculation. For a code sequence $\mathcal{C}$, in our context, $f$ can be defined as $f: \mathcal{C} \rightarrow \{0, 1\}$, or $f$ provides a binary decision whether bad speculation was observed for a given code sequence $\mathcal{C}$. We defer discussion on the generation of $\mathcal{C}$ to Section~\ref{sec:swarm_initialization} and focus on the binary decision of bad speculation here.

First, we detail on when bad speculation happens in modern Intel processors. Intel's Software Developer Manual~\cite{intel_sdm} states bad speculation occurs from three major causes - \circled{1} branch misprediction, \circled{2} microcode assists, and \circled{3} machine clears. Branch misprediction related transient execution occurs due to the need to squash micro-operations issued from the mispredicted branch and re-issue the ones from the correct branch. Likewise, microcode assists are needed when the hardware needs to rely upon microcode for execution. In such a scenario, the speculatively issued micro-operations from the micro-operation decoder are flushed from the pipeline, and are replaced by the micro-operations from the microcode sequencer (which in turn fetches these micro-operations from microcode ROM). Finally, machine clears also cause transient execution for events like memory ordering violations, self-modifying code execution, and so on.

Given this context, to implement the fitness function $f: \mathcal{C} \rightarrow \{0, 1\}$, we require an interface to capture when any of \circled{1} branch mis-prediction, \circled{2} microcode assist, or \circled{3} machine clear happens. All modern Intel systems expose a Performance Monitoring Unit~\cite{intel_perfmon} that logs different types of events occurring on the physical cores for analysis. Such performance counters are useful in understanding occurrences of bad speculation, given any code sequence $\mathcal{C}$. Table~\ref{tab:pmc} summarizes the exact performance counters we used for \textsf{Shesha}. Therefore, given a code sequence $\mathcal{C}$, the objective of the PSO will be to \textit{maximize} occurrences of bad speculation as reported by the respective performance counters.

\subsection{Disjointedly Fragmented Sub-spaces}
\label{sec:disjointedly_fragmented_sub_spaces}
We now discuss the nature of bad speculation events that we term as \emph{disjointedly fragmented} sub-spaces which allows us to achieve faster convergence for \textsf{Shesha} and efficiently scale as the number of ISE instruction pool increases.
As evident from Table~\ref{tab:pmc}, each of the performance counters pertains to bad speculation scenarios, which are \textbf{disjoint} from each other from a causal perspective. For instance, bad speculation due to memory disambiguation issues has completely different cause (incorrect speculation while resolving load/store addresses) than bad speculation due to self-modifying code execution (incorrect speculation in writing to instruction cache). Additionally, the disjoint classes of bad speculation also have effects on disjoint architectural and micro-architectural assets. For example, memory disambiguation issues will mainly cause speculation in data forwarding in load/store buffers. On the other hand, self-modifying code execution issues will affect the instruction cache. Additionally, AVX-SSE transition issues shall affect the SIMD registers and temporal buffers. Therefore, we make the following observation:

\begin{tcolorbox}[
    colback=lightgray,   
    colframe=customblack,       
    arc=2mm,             
    boxrule=1pt,          
    left=1pt,   
    right=1pt,  
    top=1pt,    
    bottom=1pt, 
]

\textit{\textbf{Equivalence Classes:} Different scenarios of bad speculation in Table~\ref{tab:pmc} are \textbf{disjoint} wrt. cause and effect (on architectural/micro-architectural assets), thereby splitting the search space of bad speculation into disjointedly, fragmented \textbf{equivalence classes}, where the cause/effect of one equivalence class are unaffected by that of other classes.}
\end{tcolorbox}

Such a definition of \emph{equivalence classes} (over set of instructions) ensures that the search space for bad speculation is split into disjointedly fragmented sub-spaces, such that the portion of the PSO's swarm working on one equivalence class need not rely upon the entirety of the PSO swarm for exploitation (cf. Section~\ref{sec:particle_swarm_optimization}). For example, without loss of generality, consider a portion of the PSO swarm working with machine clears related to self-modifying code. Then, the velocity vector computations related to swarm behaviour (i.e. involving $\mathbf{p}_{g}(i)$, cf. Section~\ref{sec:particle_swarm_optimization}) need not consider the portion of the swarm working with bad speculation due to branch mispredictions, simply because code sequences triggering branch misprediction (for instance, heavy use of loops to train the branch predictor) have no direct bearing with code sequences triggering self-modifying code (for instance, use of store operations in address ranges belonging to \texttt{.text} section of the process binary, thereby requiring write operations on the instruction cache).
Based on this description of equivalence classes, we now detail the construction of \textsf{Shesha} in the next subsection.

\subsection{Swarm Optimizer description}
\label{sec:algo_description}
With the idea of equivalence classes established, we now detail \textsf{Shesha} - an automatic analysis framework for exploring the disjointedly fragmented equivalence classes of bad speculation in ISEs. \textsf{Shesha} is divided mainly into three phases: \circleB{1} Swarm initialization, \circleB{2} Cognitive phase, and \circleB{3} Mixed phase. In phase \circleB{1}, a population pool of \texttt{N} particles is created where each particle represents a randomly sampled instruction sequence from the instruction pools of different ISEs. Then, in phase \circleB{2}, the particles are evolved to settle into sub-swarms (i.e. one swarm for one equivalence class). In this phase, since the sub-swarms are still being created, there is no exploitative behaviour (i.e. all particles evolve independently of each other). This design decision is intentional, following our discussion on equivalence classes in Section~\ref{sec:disjointedly_fragmented_sub_spaces}. Until the sub-swarms are ready, \textsf{Shesha} does not have reliable global best solutions to use for updating particle velocity (cf. Section~\ref{sec:particle_swarm_optimization} for PSO semantics). Finally, upon end of phase \circleB{2}, the sub-swarms would be ready for each of the equivalence classes, which is when phase \circleB{3} kicks in and uses both \textit{local} best and \textit{global} best solutions to update particle velocity. We now explain these phases in detail.

\subsubsection{Swarm Initialization}
\label{sec:swarm_initialization}
The first phase initializes a population of $N$ particles, where each represents an instruction sequence, sampled from the ISE instruction set. Following the semantics from Section~\ref{sec:particle_swarm_optimization}, each particle is represented by a position vector $\mathbf{x}$ in a $n$-dimensional real-valued vector space, such that the fitness function $f: \mathcal{R}^{n} \rightarrow \mathcal{R}$ can be evaluated upon the same. We describe now how we map a code sequence to a real-valued $n$-dimensional vector.

To achieve this, we leverage a machine-readable \texttt{x86} ISE list from uops.info~\cite{Abel19a} and use it to construct the ISE instruction set pool, further sub-divided into about $90$ ISE types, like \texttt{AES}, \texttt{AVX}, \texttt{AVX2}, \texttt{AVX512EVEX}, \texttt{AVX512VEX}, \texttt{AVXAES}, \texttt{BMI1}, \texttt{BMI2}, \texttt{SSE}, \texttt{SSE2}, \texttt{SSE3}, \texttt{SSE4}, \texttt{X87} to name a few. \textsf{Shesha} allows the user to choose which ISEs to be explored. To generate a single particle, \textsf{Shesha} randomly samples $n$ instructions from the selected instruction pool as well as randomly selects operands (for instance, which registers to use out of the entire register bank). Following the description of fitness function discussed in~\ref{sec:fitness_function}, this ordered set of $n$ sampled instructions and their operands corresponds to the instruction sequence $\mathcal{C}$. In our experiment, we set the values of $N$ and $n$ as $50$ and $10$, respectively (without loss of generality).

Next, we derive a position vector $\mathbf{x}$ in the $n$-dimensional real vector space from $\mathcal{C}$. To do so, we first assign a unique real-value \texttt{opcode} to each instruction in the entire set of ISEs considered by {\sf Shesha}. Likewise, the operand description of the instruction \texttt{opcode} is also assigned a uniquely identifiable real-value \texttt{operand\_identifier\_bitstring}. The \texttt{opcode} and the \texttt{operand\_identifier\_bitstring} can be combined to uniquely identify any instruction amongst all possible instructions in the ISEs under consideration. We note here that \texttt{operand\_identifier\_bitstring} depends simply on the operands used, and \textit{not} on the actual data held by the operands when the instruction is executed. Finally, the \textit{combined} representation of an instruction is uniquely identifiable by \texttt{(opcode << i) | operand\_identifier\_bitstring}, where $i$ is the number of bits needed to uniquely identify the operands. We explain the mapping using an example. Assume the sampled instruction is \texttt{VPXOR xmm1, xmm2, xmm3}, where \texttt{VPXOR} is the AVX flavor of Logical XOR on SIMD registers \texttt{xmm1}, \texttt{xmm2}, \texttt{xmm3}. Now \texttt{VPXOR} shall have a unique \texttt{opcode} identifying this instruction, while each of the operand registers can be uniquely identified by a $4$ bit value (since there are $16$ unique \texttt{xmm} registers). Hence, the real-valued unique identifier for this instruction shall be \texttt{(vpxor\_opcode << 12) | (1 << 8) | (2 << 4) | 3}, assuming \texttt{xmm{k}} register to be uniquely identifiable by the $4$ bit integer $k$. Repeating the same procedure for all sampled instructions creates $n$ distinct real values, which contribute to the $n$-dimensional position vector $\mathbf{x}$. 
At the end of this phase, we have the set of position vectors $\mathbb{X} = \{\mathbf{x}_{1}, \mathbf{x}_{2}, ..., \mathbf{x}_{N}\}$. Next, we describe how to compute velocity vectors in order to mutate these position vectors.

\begin{table*}[!pt]
\centering
\caption{Leakage Variants Discovered by {\sf Shesha}}.
\label{tab:leakage_variants}
\begin{threeparttable}
\resizebox{\textwidth}{!}{%
\begin{tabular}{|l|l|l|l|l|l|l|l|}
\hline
\multicolumn{1}{|c|}{\textbf{Cases}} &
  \multicolumn{1}{c|}{\textbf{Operations}} &
  \multicolumn{1}{c|}{\textbf{\begin{tabular}[c]{@{}c@{}}Register\\ Dependency\end{tabular}}} &
  \multicolumn{1}{c|}{\textbf{\begin{tabular}[c]{@{}c@{}}Equivalence\\ Class\end{tabular}}} &
  \multicolumn{1}{c|}{\textbf{Assets Affected}} &
  \multicolumn{1}{c|}{\textbf{Generation}} &
  \multicolumn{1}{c|}{\textbf{\begin{tabular}[c]{@{}c@{}}Window\\ Size\end{tabular}}} &
  \multicolumn{1}{c|}{\textbf{Attack Type}} \\ \hline
\circled{1} &
  SIMD-Vector intermixing &
  $\times$ &
  SMC, SAM &
  \begin{tabular}[c]{@{}l@{}}I-Cache, \\ SIMD registers\end{tabular} &
  \begin{tabular}[c]{@{}l@{}}12G \checkmark, 11G $\times$\\ 4X \checkmark, 3X $\times$  \end{tabular} & $17$ ($1$ \texttt{rep})
   &
  Leak \\ \hline
\circled{2} &
  \begin{tabular}[c]{@{}l@{}}Single and Double precision\\ intermixing\end{tabular} &
  \checkmark &
  HW, SMC, MO &
  \begin{tabular}[c]{@{}l@{}}I-Cache, precision\\ conversion hardware,\\ SIMD buffer\end{tabular} & \begin{tabular}[c]{@{}l@{}}12G \checkmark, 11G $\times$\\ 4X \checkmark, 3X $\times$  \end{tabular}
   & $20$ ($100$ \texttt{rep})
   &
  Leak \\ \hline
\circled{3} &
  Fused Multiply and Addition &
  $\times$ &
  FPA &
  FPU &
  \begin{tabular}[c]{@{}l@{}}12G \checkmark, 11G \checkmark\\ 4X \checkmark, 3X \checkmark   \end{tabular} & $12$ ($32$ \texttt{rep})
   &
  \begin{tabular}[c]{@{}l@{}}Leak,\\ Injection\end{tabular} \\ \hline
\circled{4} &
  \begin{tabular}[c]{@{}l@{}}SSE-AES using Denormal\\ numbers\end{tabular} &
  \checkmark &
  FPA, SMC &
  FPU, AES hardware & \begin{tabular}[c]{@{}l@{}}12G \checkmark, 11G \checkmark\\ 4X \checkmark, 3X \checkmark   \end{tabular}
   & $12$ ($32$ \texttt{rep})
   &
  \begin{tabular}[c]{@{}l@{}}Leak,\\ Injection\end{tabular} \\ \hline
\end{tabular}%
}
\begin{tablenotes}
      \scriptsize
      \item {\sf SMC}: Self-Modifying Code; \quad {\sf SAM}: SSE-AVX Mix; \quad {\sf HW}: Hardware Assist;  \quad {\sf FPA}: Floating Point Assist;  \quad {\sf MO}: Memory Ordering; \quad {\sf rep}: No. of repetitions
      \item {\sf Leak}: Transient Leakage through Flush+Reload;  \quad {\sf Injection}: Floating Point Value Injection; \quad \checkmark: Required or Available; \quad $\times$: Not required or Not available
      \item {\sf FPU}: Floating point unit; \quad {\sf 12G}: Alder Lake; \quad {\sf 11G}: Comet Lake; \quad {\sf 4X}: Sapphire Rapids; \quad {\sf 3X}: Ice Lake
    \end{tablenotes}
\end{threeparttable}
\end{table*}

\subsubsection{Cognitive phase}
\label{sec:cognitive_phase}

In this phase, we allow for the \emph{compartmentalization of the swarm population} into the disjointedly fragmented equivalence classes. Since the initialization phase has already computed the set of position vectors $\mathbb{X} = \{\mathbf{x}_{1}, \mathbf{x}_{2}, ..., \mathbf{x}_{N}\}$, we focus on generation of velocity vectors that allow generation of sub-swarms (i.e. one sub-swarm for each equivalence class). In order to do this, we must prevent any exploitation/swarm behaviour (cf. Section~\ref{sec:particle_swarm_optimization}) from kicking in the cognitive phase. For this case, we choose $\alpha = 1$, cognitive acceleration coefficient $\beta$ as $0.4$, and social acceleration coefficient $\gamma$ as $0$ (to cancel out any swarm behaviour). This forces using the local best solutions $\mathbf{p}$ with probability $\beta$. In this phase, we rely upon two kinds of mutations to the velocity vector:

\begin{enumerate}
    \item \textbf{Modification to operands}: Given a particle position vector $\mathbf{x}_{j}: 1 \le j \le N$, we sample a random dimension $d: 1 \le d \le n$ and mutate the operands with probability $\beta$. Following the semantics of~\ref{sec:swarm_initialization}, assume $x_{j}^{d} =$ \texttt{(opcode << i) | operand\_identifier\_bitstring}), where | represents Logical OR.  Thereby, the new position vector is evaluated as $\mathbf{x}_{j}^\prime = \{x_{j}^{1}, x_{j}^{2}, ..., x_{j}^{d-1}, (x_{j}^{d} \, \wedge \,$ \texttt{(1 << i)}) \texttt{<< i | new\_operand\_identifier\_bitstring}, $x_{j}^{d+1}, ..., x_{j}^{n}\}$. Here, the newer operands that are sampled are converted to \texttt{new\_operand\_identifier\_bitstring} using the same semantics as in Section~\ref{sec:swarm_initialization}. To follow upon the same example of \texttt{VPXOR xmm1, xmm2, xmm3} having representation \texttt{(vpxor\_opcode << 12) | (1 << 8) | (2 << 4) | 3}) in the position vector, assume the operand mutation converted this instruction to \texttt{VPXOR xmm5, xmm13, xmm15}. Then the modified representation of the aforementioned instruction in the position vector can be given by \texttt{(vpxor\_opcode << 12) | (5 << 8) | (13 << 4) | 15}).

    \item \textbf{Modification to instruction}: Like operand modifications, here also $x_{d} =$ \texttt{(opcode << i) | operand\_identifier\_bitstring}) is mutated as $x_{d}^{\prime} = $ \texttt{(new\_opcode << i) | operand\_identifier\_bitstring}) with probability $\beta$ (cognitive acceleration coefficient), where \texttt{new\_opcode} is a newly sampled instruction from the ISE set. Again taking the same example, suppose \texttt{VPXOR xmm1, xmm2, xmm3} is mutated to \texttt{VPAND xmm1, xmm2, xmm3}, then \texttt{(vpxor\_opcode << 12) | (1 << 8) | (2 << 4) | 3}) shall be mutated to \texttt{(vpand\_opcode << 12) | (1 << 8) | (2 << 4) | 3}).
\end{enumerate}

Finally, the mutated position vector is evaluated upon the fitness function as defined in Section~\ref{sec:fitness_function} to understand if the mutated particle discovers some newer avenues of bad speculation. As iterations of the Cognitive phase increase, the entire population of particles converges upon one of the several disjointedly fragmented equivalence classes. Thereby, at the end of the Cognitive phase, the sub-swarms belonging to each equivalence class $\mathcal{S}_{1}, \mathcal{S}_{2}, ..., \mathcal{S}_{9}$ are output to phase \circleB{2} (where each $S_{i}$ belongs to one equivalence class as in Table~\ref{tab:pmc}). Concretely, each particle in the initialized swarm belongs to exactly one of these equivalence classes.

\subsubsection{Mixed phase}
\label{sec:mixed_phase}

In this phase, we enable swarm behaviour onto the discrete sub-swarms $\mathcal{S}_{1}, \mathcal{S}_{2}, ..., \mathcal{S}_{9}$ output from the Cognitive phase. Concretely, the Mixed phase makes the following changes to Cognitive phase settings: \circled{1} reduced rate of exploration, \circled{2} increased rate of exploitation from the sub-swarm $\mathcal{S}_{i}$ to which the particle belongs to. To achieve this, we set $\alpha = 1$, cognitive acceleration coefficient $\beta = 0.1$ (as opposed to $0.4$ in the Cognitive phase), and the social acceleration coefficient $\gamma = 0.4$ (as opposed to $0$ in the Cognitive phase). This setting still allows exploration using the local best solutions, albeit with reduced probability $\beta$, while also starting to use the global best solutions found in the sub-swarm $\mathcal{S}_{i}$ to which the particle belongs. In this phase, we rely upon three kinds of mutations to the velocity vector:

\begin{enumerate}   \itemsep0em
    \item \textbf{Modification to operands}: Same as Cognitive phase, but with reduced probability $\beta = 0.1$ instead of $0.4$.
    
    \item \textbf{Modifications to instruction}: For each particle position vector $\mathbf{x}_{j}$ belonging to some sub-swarm $\mathcal{S}_{k}$, sample two random dimensions $d_{1}, d_{2}: 1 \le d_{1}, d_{2} \le n$ and mutate $x_{j}^{d_{1}}$ as in Cognitive phase with probability $\beta = 0.1$. For dimension $d_{2}$, with probability $\gamma = 0.4$, replace the $d_{2}$-th dimension of $\mathbf{x}_{j}$ with the $d_{2}$-th dimension of the leader of sub-swarm $S_{k}$. Here, \textit{leader} of the sub-swarm $\mathcal{S}_{k}$ is defined as the position vector in sub-swarm $S_{k}$ with maximum fitness as described in Section~\ref{sec:fitness_function} (i.e. the instruction sequence with maximum bad speculation in the equivalence class $\mathcal{S}_{k}$).

    \item \textbf{Dimensionality Reduction}: For a given position vector $\mathbf{x}_{j}: 1 \le j \le N$ belonging to sub-swarm $S_{k}$ and defined as $\mathbf{x}_{j} = \{x_{j}^{1}, x_{j}^{2}, x_{j}^{3}, ..., x_{j}^{n}\}$, remove a randomly sampled dimension $d: 1 \le d \le n$ with probability $\beta = 0.1$. One might recall that each element in $\mathbf{x}_{j}$ is representative of an instruction. Hence, informally, we can treat the real-valued vector $\mathbf{x}_{j}$ as an instruction sequence $\mathcal{C}$ comprising of $n$ instructions. Therefore, dimensionality reduction aims to \textit{reduce} this instruction sequence so as to generate smaller and focused sequences, while still maximizing the occurrence of bad speculation.
\end{enumerate}

At the end of this phase, \textsf{Shesha} outputs the population position vectors with maximized fitness (i.e. maximum occurrence of bad speculation) in the respective equivalence classes. We note that dimensionality reduction, in particular, ensures that the final output of \textsf{Shesha} is able to drop instructions from the generated instruction sequence that may not contribute to bad speculation. In other words, because of dimensionality reduction, the final output of {\sf Shesha} is a \textit{minimal} reproduced instruction sequence responsible for the relevant bad speculation scenarios detailed in Table~\ref{tab:pmc}.

\section{Uncovered Transient Execution Paths}
\label{sec:transient_execution_paths}

We ran a test campaign of $24$ hours, in which most observations were obtained within the first $8$ hours (refer to Algo.~\ref{sec:algo_description_Shesha} in Appendix for description of the algorithm). Since we rely upon Intel Performance Monitoring Unit (PMU), which captures the \textit{events} occurring in the system with certainty, the false positive rate of {\sf Shesha} is zero. In this section, we detail the novel transient execution paths uncovered by \textsf{Shesha} as a result of the testing campaign. We also reverse engineer the root-cause of these transient execution paths and detail their effect on the overall behaviour of the system under test. 

\begin{enumerate}

    \item \textbf{Hardware assists for precision conversion}: All floating point numbers can have two precision levels: single or double. With dedicated instructions for single precision and double precision separately, the same set of registers (i.e. $\{$\texttt{xmm0, xmm1, ...}$\}$, $\{$\texttt{ymm0, ymm1, ...}$\}$ and $\{$\texttt{zmm0, zmm1, ...}$\}$) need to be interpreted as single precision or double precision based on the instruction operand. In this context, we uncover a \textbf{hardware assist} that kicks in case of data dependencies between instructions that intermix $\{$\texttt{xmm0, xmm1, ...}$\}$, $\{$\texttt{ymm0, ymm1, ...}$\}$ and $\{$\texttt{zmm0, zmm1, ...}$\}$ in different precision modes, where the data in the SIMD register bank needs to be interconverted between single and double precision modes.

    \item \textbf{Self-modifying code execution due to intermixing precision}: Likewise, intermixing different representations of $\{$\texttt{xmm0, xmm1, ...}$\}$, $\{$\texttt{ymm0, ymm1, ...}$\}$ and $\{$\texttt{zmm0, zmm1, ...}$\}$ wrt. precision also causes \textbf{machine clears} involving updates to the instruction cache, enabling yet another transient execution path.

    \item \textbf{Memory ordering issues due to intermixing precision}: We also uncover that intermixing instructions with single/double precision causes \textbf{memory ordering based machine clear}, uncovering a new transient execution path not discussed prior in literature.

    \item \textbf{Intermixing SIMD-Vector instructions}. A wide variety of ISEs like \texttt{AVX}, \texttt{AVX2}, \texttt{AVX512EVEX}, \texttt{AVX512VEX}, \texttt{SSE}, \texttt{SSE2}, \texttt{SSE3}, \texttt{SSE4}, \texttt{SSE4a}, and \texttt{SSSE3} share the SIMD register banks $\{$\texttt{xmm0, xmm1, ...}$\}$, $\{$\texttt{ymm0, ymm1, ...}$\}$ and $\{$\texttt{zmm0, zmm1, ...}$\}$. As such, transitioning from SIMD to Vector instructions or vice-versa incurs a machine clear attributed to self-modifying code. We note that although older Intel generation processors like Skylake had transition penalties related to intermixing of SIMD-Vector instructions~\cite{intel_sdm}, recent Intel processors up to Comet Lake have no such penalties~\cite{ragab2021rage}. However, as we uncover, the $12$-th generation client processor (Alder Lake) and $4$-th generation Xeon processor (Sapphire Rapids) show transient execution due to an involved \textbf{machine clear due to self-modifying code}.

    \item \textbf{FP assist due to Denormal arithmetic}: We also uncover newer avenues of \textbf{Floating Point assists} in the Fused-Multiply-Add instruction family as well during transitions between AES-NI and SSE instructions. Although Floating Point assists have been demonstrated before~\cite{ragab2021rage}, our tool and analysis show that Floating Point assists have a wider scope than simple SIMD arithmetic (as previously discussed in literature).
\end{enumerate}

\section{\textsf{Shesha} Design Choices}
\label{sec:ablation_study}
We now explain the choice of parameters by comparing the role of Particle behaviour and Swarm behaviour in {\sf Shesha}. To do so, we define $6$ variants of {\sf Shesha} wrt. hyperparameters $\beta$ (which controls Particle behaviour) and $\gamma$ (which controls Swarm behaviour). To compare between the variants, we define two metrics: \textcircled{1} time taken to find the \textit{first} SIMD-Vector speculation, and \textcircled{2} time taken to find the \textit{first} precision intermixing speculation. The reason for this (as detailed in subsequent sections) is because SIMD-Vector speculation is the easiest of the four (cf. Table~\ref{tab:leakage_variants}) to discover. This is because of the large number of instructions under these ISEs, as well as no special requirement (like read-after-write dependencies, or specific floating-point arithmetic) needed to trigger this speculation. Precision intermixing, however, is on the other extreme, and is the hardest of the four to uncover. This is because (as we detail in Section~\ref{sec:single_double_precision}) such speculation requires closely tied instructions that handle single/double precision respectively, and have read-after-write dependency.

Variants {\sf Shesha}$_{\{\beta = 1, \gamma = 0\}}$ and {\sf Shesha}$_{\{\beta = 0, \gamma = 1\}}$ are essentially equivalent to random sampling, with different sources. The former prioritizes mutating the chosen instruction within the particle itself by replacing it with a random instruction with probability $1$. The latter also works to similar goal, but replaces the chosen instruction with another instruction from other particle in the swarm. Then, variants {\sf Shesha}$_{\{\beta = 0.4, \gamma = 0\}}$ and  {\sf Shesha}$_{\{\beta = 0.1, \gamma = 0\}}$ test the extent of particle behaviour by changing the probability $\beta$ of replacing a chosen instruction with a randomly sampled instruction. Finally, variants {\sf Shesha}$_{\{\beta = 0.1, \gamma = 0.1\}}$ and {\sf Shesha}$_{\{\beta = 0.1, \gamma = 0.4\}}$ test the extent of swarm behaviour by changing the probability $\gamma$ of replacing a chosen instruction with an instruction from another particle in the swarm. We summarize these in Table~\ref{tab:Shesha_variant}.

First, we note that {\sf Shesha}$_{\{\beta = 0.1, \gamma = 0\}}$ is inherently slow in both aspects we consider here: it adopts a \textit{random} sampling strategy (by setting $\gamma = 0$), and performs such sampling at a reduced rate of $0.1$. The best performance (wrt. Precision Intermixing) is given by {\sf Shesha}$_{\{\beta = 0.1, \gamma = 0.4\}}$, which allots a slightly higher probability of swarm behaviour (through $\gamma = 0.4$) than other variants. However, it is the variant {\sf Shesha}$_{\{\beta = 0.4, \gamma = 0\}}$ that outperforms all wrt. SIMD-Vector intermixing, possibly due to balancing the aggressiveness with which to replace instructions. Interestingly, note that variants equivalent to random sampling (i.e. {\sf Shesha}$_{\{\beta = 1, \gamma = 0\}}$ and {\sf Shesha}$_{\{\beta = 0, \gamma = 1\}}$) perform fairly poorly wrt. SIMD-Vector intermixing (possibly due to very aggressive mutation strategy).

As a result, in our final deployment of {\sf Shesha}, we adopt a mixture of variant {\sf Shesha}$_{\{\beta = 0.4, \gamma = 0\}}$ and variant {\sf Shesha}$_{\{\beta = 0.1, \gamma = 0.4\}}$. The first variant forms the Cognitive phase (cf. Section~\ref{sec:cognitive_phase}), while the second variant forms the Mixed phase (cf. Section~\ref{sec:mixed_phase}). Finally, we note that variant {\sf Shesha}$_{\{\beta = 1, \gamma = 0\}}$ \textit{simulates} other \textit{fuzzing} (essentially relying upon a random sampling strategy) based tools like~\cite{xiao2019speechminer,oleksenko2022revizor,moghimi2020medusa,oleksenko2023hide} had they supported ISEs like AVX/SSE/FMA and so on. For comparison however, we report the raw data from these works, albeit with different objectives over mostly x86 ISA. For instance, wrt. the speculative paths and ISA explored by~\cite{oleksenko2022revizor}, it is able to detect violations in about $5$ minutes (which is comparable to {\sf Shesha}$_{\{\beta = 0.4, \gamma = 0\}}$'s detection time for SIMD-Vector transient path detection). Likewise, in~\cite{moghimi2020medusa}, the entire fuzzing campaign lasts about $20+$ CPU hours. On similar lines, the detection in~\cite{oleksenko2023hide} ranges from a minute to about an hour based on the type of speculation targeted~\footnote{\cite{xiao2019speechminer} does not report convergence time.}. One must note that the objectives of each of these tools (as presented in their corresponding papers) were different, and thus, they explore different subsets of ISAs. \textsf{Shesha} works with comparatively larger ISA set and thus explores wider range of instruction combinations and stimulates variety of specialized execution units and buffers.

\begin{table}[!t]
      \centering
  \caption{Time taken to uncover speculation by the variant.}
  \label{tab:Shesha_variant}
  \resizebox{0.9\linewidth}{!}{
  \begin{tabular}{|c|c|c|}
    \hline
    \textbf{Variant}  & \textbf{SIMD-Vector} & \textbf{Precision Intermixing}  \\
    \hline
    {\sf Shesha}$_{\{\beta = 1, \gamma = 0\}}$ & $0.5$ hour & $14$ hours\\
    \hline
    {\sf Shesha}$_{\{\beta = 0.4, \gamma = 0\}}$ & $0.01$ hour & $8$ hours \\
    \hline
    {\sf Shesha}$_{\{\beta = 0.1, \gamma = 0\}}$ & $1$ hour & $\approx 24$ hours \\
    \hline
    {\sf Shesha}$_{\{\beta = 0.1, \gamma = 0.1\}}$ & $0.6$ hour & $4$ hours \\
    \hline
    {\sf Shesha}$_{\{\beta = 0.1, \gamma = 0.4\}}$ & $0.03$ hour & $2$ hours \\
    \hline
    {\sf Shesha}$_{\{\beta = 0, \gamma = 1\}}$ & $0.5$ hour & $9$ hours \\
    \hline
\end{tabular}}
\end{table}

\section{Root-Cause Analysis of Discovered Transient Execution Paths}~\label{sec:root_cause}
We now provide root-cause analysis for the two transition-based leakages discovered by \textsf{Shesha}, namely, SIMD-Vector transition and single-double precision transitions.

\color{black}

\subsection{SIMD-Vector Transition based SMC}
\label{sec:simd_vector_transition_smc}
SIMD and Vector instruction extensions like \texttt{AVX}, \texttt{AVX2}, \texttt{AVX512EVEX}, \texttt{AVX512VEX}, \texttt{SSE}, \texttt{SSE2}, \texttt{SSE3}, \texttt{SSE4}, \texttt{SSE4a}, and \texttt{SSSE3} are central to several performance gains experienced in modern Intel processors. These ISEs allow greater than $64$ bit registers holding more than one data point, to be operated in parallel by the same instruction (hence the name SIMD type ISEs). To do so, modern IA-64 architectures define a register bank with sixteen $512$-bit wide registers. Interestingly, ISEs like \texttt{AVX2}/\texttt{SSE} allow \textit{differing} representations of the \textit{same} hardware register in two dimensions: \textcircled{1} width $w$ of the register, and \textcircled{2} the SIMD representation of the data contained in that width $w$. To elaborate on dimension \textcircled{1}, the \textit{same} hardware registers can be interpreted as $512$-bit wide registers \texttt{zmm0, zmm1, zmm2, ..., zmm15} (by ISEs like \texttt{AVX512EVEX}, \texttt{AVX512VEX}), as $256$-bit wide registers \texttt{ymm0, ymm1, ymm2, ..., ymm15} (by ISEs like \texttt{AVX2}), and as $128$-bit wide regisers \texttt{xmm0, xmm1, xmm2, ..., xmm15}. Likewise, regarding dimension \textcircled{2}, the same data in any \texttt{zmm}/\texttt{ymm}/\texttt{xmm} register can be interpreted as differently packed integer/floating-point value. This discussion raises a concerning question: \textit{Given a processor allowing executions of ISEs requiring different representations of the same hardware, how are transitions between such ISEs handled}? According to the Intel Software Developer Manual~\cite{intel_sdm}, the exact implementation to handle such transitions depends upon the exact micro-architecture under question. We defer the details of these implementations and our reverse-engineering approach to Appendix~\ref{appx:simd_vector_transient}. We make the following observation as the root cause of bad speculation (as in Table~\ref{tab:leakage_variants}) encountered in modern Intel processors when transitioning from SIMD to Vector instructions (and vice-versa). 

\begin{tcolorbox}[
    colback=lightgray,   
    colframe=customblack,       
    arc=2mm,             
    boxrule=1pt,          
    left=1pt,   
    right=1pt,  
    top=1pt,    
    bottom=1pt, 
]

\emph{\textbf{Observation}. By deciding to remove the Finite State Machine based register dependency analysis and insertion of blend operations into the issue state of the pipeline, micro-optimizations for Alder Lake and Sapphire Rapids handle SIMD-Vector transitions by incurring a self-modifying code execution that inserts the blend operations directly into the instruction cache, therefore causing speculation.}
\end{tcolorbox}

\subsection{Precision based Transient Execution Paths}
\label{sec:single_double_precision}
Modern ISEs usually work upon a register bank of $16$ registers of width $512$ bits, and the same data on these registers can be interpreted in different ways. While it is straightforward to switch between integer representations, switching between single-precision and double-precision representations is an involved process. Through the course of our testing campaign, \textsf{Shesha} uncovered a previously unknown source of transient execution paths: \textit{non-explicit switching between single and double precision floating-point numbers in a Read-after-Write (RAW) data dependency}. Concretely, a sequence of two instructions linked by a RAW dependency causes novel transient execution paths on Intel CPUs if they require contrasting representations (single precision vs double precision) of the same floating-point number. Interestingly, three independent, completely novel transient execution paths are uncovered by \textsf{Shesha} as a result of single/double precision intermixing with RAW data dependency- \circled{1} hardware assist (HW), \circled{2} self-modifying code (SMC) based machine clear, and \circled{3} memory ordering (MO) violation based machine clear. For all subsequent discussions, we consider the following abstract instruction sequence (\texttt{SP}: single precision instruction, \texttt{DP}: double precision instruction):
\begin{enumerate}   \itemsep0em
    \item \texttt{SP xmm0, xmm1, xmm2} followed by \texttt{DP xmm3, xmm4, xmm0} \, \, \textbf{OR}
    \item \texttt{DP xmm0, xmm1, xmm2} followed by \texttt{SP xmm3, xmm4, xmm0}
\end{enumerate}


Precisely, for the discussed transient paths to kick in, the requirement is a set of at least two instructions with a Read-after-Write (RAW) dependency connecting them (using the \texttt{xmm0} register in the above examples), without occurrence of an explicit conversion instruction (like \texttt{CVTPD2PS} or \texttt{CVTPS2PD}) in between. We note here that Write-after-Write (WAW) or Write-after-Read (WAR) dependencies do \textit{not} trigger any of the following transient paths, simply because there is no need for any implicit precision conversion as the second instruction is \textit{writing} to the dependent register and not reading from it. Writes from a single-precision instruction to a register holding a double-precision value are architecturally simply \textit{overwriting} the registers, and thus incur no transient executions. We defer the detailed discussion on each of these transient execution paths and our reverse-engineering approach to Appendix~\ref{appx:precision_transient}. We make the following observation as the root cause of all the bad speculation events when transitioning between single and double precision.

\begin{tcolorbox}[
    colback=lightgray,   
    colframe=customblack,       
    arc=2mm,             
    boxrule=1pt,          
    left=1pt,   
    right=1pt,  
    top=1pt,    
    bottom=1pt, 
]

\emph{\textbf{Observation}. Owing to differing representations of same data in \texttt{xmm}/\texttt{ymm}/\texttt{zmm} wrt. single and double precision, a read-after-write (RAW) dependency without \texttt{CVTPD2PS/CVTPS2PD} incurs a hardware assist to perform implicit conversion, hence triggering aggressive speculation. Moreover, accompanying machine clears related to self-modifying code and memory ordering are also involved, thus contributing to speculation.}
\end{tcolorbox}



\color{black}
\section{Attack Building Blocks}
\label{sec:attack_building_blocks}

\color{black}

\subsection{Transient Execution Window}
\label{sec:transient_execution_window}
Transient execution paths guarantee flushing of the in-flight micro-operations from the pipeline and re-steer the execution on the correct path. As such, the \textit{width} of the transient window for any given transient execution path determines the number of micro-operations an attacker can issue on the transient path before the rollback of the instructions. Practically, a larger transient window allows for executing complex attack vectors in the transient path. Architecturally, no effect of any transient micro-operation is visible. Therefore, we need to rely upon micro-architectural effects to measure the transient window. In our case, we compute the transient window by counting the number of transient loads executed before the pipeline rollback. When no bad speculation (discovered by {\sf Shesha}) occurs, we observe no activity in the transient path, implying absence of misprediction-based transient execution. Our results are summarized in Table~\ref{tab:leakage_variants}. In comparison to the machine clears presented in~\cite{ragab2021rage}, we conclude that the SIMD-Vector transition-based SMC exhibits a larger transient window than all other variants of transient executions through machine clears or assists (both uncovered here as well as in~\cite{ragab2021rage}). This observation directly ties in with the number of \textit{repetitions} (for warming up the execution pipeline) an attacker needs to do on the transient path before the actual attack can be mounted (cf. Section~\ref{sec:rep_transient_window}). As evident, due to the largest transient window with minimum number of repetitions, SIMD-Vector transition-based SMC provides a very practical transient execution path (with effectively no repetitions required), thereby becoming an actual choice for an adversary to mount in-the-wild attacks.

\noindent \textbf{Enlarging the transient window.} 
\label{sec:rep_transient_window}
As different transient execution paths have differing transient windows, not all of them would be directly suited to leak reliably. To circumvent this, a usual method is to warm up the pipeline before the actual attack; this can be done solely on the attacker's end. This can be done by several repetitions (\texttt{rep}) of the same transient operations to maximize the transient window. To determine the extent of pipeline warm-up needed to leak data, we increase the repetitions of warm-up until we observe leakage from the Flush+Reload~\cite{yarom2014flush+} covert channel. In Table~\ref{tab:reps}, we provide a comparative analysis of the number of repetitions needed to warm up the pipeline. Lesser the number of required repetitions \texttt{rep}, more likely it is to find in-the-wild occurrence of the uncovered transient paths in in-the-wild code bases~\footnote{We could not test the pipeline warmup and leakage rates for ~\cite{xiao2019speechminer} and \cite{stecklina2018lazyfp} because of absence of Intel TSX on our test platforms.}.

\subsection{SMT Covert Channel}
\label{sec:smt_covert_channel}
We demonstrate a covert channel using contention in the affected hardware assets. Two processes are pinned to two logical cores of the same physical core. Hence, we assume Simultaneous Multi-Threading (SMT) or HyperThreading to be enabled for the covert channel construction to work since the affected assets identified (in Table~\ref{tab:leakage_variants}) are shared in an SMT setting. The construction of the covert channel relies upon two actors: a \textit{sender} and a \textit{receiver}, executing on the same physical core (cf. Algo.~\ref{algo:covert_channel_sender} and Algo.~\ref{algo:covert_channel_receiver} in Appendix). The receiver \textit{times} the execution of a code sequence $\mathcal{C}$ ($\mathcal{C}$ has no possibility of any transient execution). In contrast, depending on whether the sender has to send bit $1$ or $0$, it performs an additional execution involving transient execution. The presence or absence of transient execution causes contention in the concerned hardware assets, leading to a perceivable timing difference on the receiver side. Such a covert channel instantiated with SIMD-Vector transition transient executions has a bandwidth of $2$ KB/s with an accuracy of $87\%$.

\begin{table}[!t]
      \centering
  \caption{Comparative analysis of the extent of pipeline warmup needed by different transient paths. Leakage rate is computed as the ratio of number of transient loads to the \texttt{rep} needed to observe visible differences in Flush+Reload~\cite{yarom2014flush+}.}
  \label{tab:reps}
  \resizebox{0.9\linewidth}{!}{
  \begin{tabular}{|c|c|c|}
    \hline
    {\bf Transient Path}  & {\bf Required} \texttt{rep} & {\bf Leakage Rate}  \\
    \hline
    FP assist \cite{ragab2021rage}  & $32$ & $0.375$ \\
    SMC assist \cite{ragab2021rage} & $20$ &  $0.3125$ \\
    Snoop MO \cite{ragab2021rage}  & $150$ &  $0.4$ \\
    MD \cite{ragab2021rage} & $32$ & $0.53125$ \\
    Page Fault~\cite{moghimi2023downfall,moghimi2020medusa} & $1$ & $0.1$ \\
    SIMD-Vector (\textbf{Our work}) & $1$ & $17$ \\
    D2S/S2D (\textbf{Our work}) & $100$ & $0.2$ \\
    FMA/AES (\textbf{Our work}) & $28$ & $0.42$ \\
    \hline
\end{tabular}}
\end{table}

\subsection{Load Value Injection}
\label{sec:lvi}

In addition to the \textbf{Leak} attack type summarized in Table~\ref{tab:leakage_variants} and detailed in Section~\ref{sec:rep_transient_window}, the Floating Point operations relying upon an assist to carry out the denormalization operation forward incorrect \textit{data} to subsequent transient instructions, before denormalization finishes. This scenario is a floating point flavor of classic load value injection (LVI) in transient executions~\cite{ragab2021rage}. However, unlike Section~\ref{sec:rep_transient_window}, the exact transient value injected into the transient operations is not architecturally visible. Hence, the adversary needs to mount LVI in two phases: \textcircled{1} offline phase where the adversary fuzzes the transient operation using different inputs and constructs the transient output, and \textcircled{2} online phase where the adversary uses the specially crafted inputs from \textcircled{1} that give a transient output of adversarial interest. On comparing the architectural result of the computation against the constructed LVI output, we observed both to be different, implying a successful transient value injection attack.

\subsection{FP Assists due to FMA instructions}
\label{sec:fp_assists_fma}

According to IEEE-754 standard, any floating point representation of a number consists of three parts: \textcircled{1} sign bit, \textcircled{2} biased exponent $e$, and \textcircled{3} mantissa $m$. The standard also defines \textit{denormalization}: a choice to trade precision for wider range of floating point numbers that can be represented. Concretely, if $e = 0$ and $m \ne 0$, denormalization can allow for a gradual underflow by appending enough leading zeroes to the mantissa until the minimal exponent is achieved.

As stated in~\cite{ragab2021rage}, denormalization requires a Floating Point assist (denoted by \texttt{ASSISTS.FP} in Table~\ref{tab:pmc}), which kicks in micro-operations to \textit{gradually underflow} a floating point number close to $0$.  This causes a flush of the current in-flight micro-operations from the execution pipeline, thereby creating a transient window where non-\textit{denormalized} floating-point values are forwarded to subsequent instructions speculatively, resulting in a transient window of executions on incorrect data, while the floating-point assist is \textit{denormalizing} the concerned value. In~\cite{ragab2021rage}, this transient window is used to construct a flavor of load-value injection exploits on the \texttt{SSE2} instruction extension. Through {\sf Shesha} however, we uncover existence of this transient execution path in other ISEs as well~\footnote{{\sf Shesha} also uncovers previously unreported combinations of AES and SSE instruction sequences (like \texttt{AESDECLAST} and \texttt{MULSS} sharing a Read-after-Write register dependency) to also incur Floating-Point assists.}, most notably in Fused-Multiply-Add (FMA) instructions like \texttt{VFMADD132SD}, \texttt{VFMADD132PD}, or exponentiation instructions like \texttt{VSCALEFSS}. 


\section{Exploiting Leakage from FMA}
\label{sec:fma_leaking}

An interesting finding by \textsf{Shesha} is the transient nature of the Fused Multiply-Add (FMA) instruction set extension, which warrants a deeper investigation. Although the transient nature of AVX instructions is well-known~\cite{moghimi2023downfall, ragab2021rage}, the security implications of FMA (and its variants) instructions have not been studied in literature before. In this section, we develop on the findings of Section~\ref{sec:fp_assists_fma}, and focus on the exploitability of previously unreported speculation in FMA execution. 


\subsection{Fused Multiply-Add Execution Unit}
\label{sec:fma_details}

FMA instructions have been traditionally introduced to accelerate arithmetic capabilities of Intel processors. 
Such instruction extensions are capable of performing \textit{fused} arithmetic operations in one instruction execution. The classic FMA family of instructions can support operations like: \textcircled{1} fused multiply-add, \textcircled{2} fused multiply-subtract, \textcircled{3} fused multiply add/subtract interleave, \textcircled{4} signed-reversed multiply on fused multiply-add and multiply-subtract. Moreover, several new ``FMA-like'' instruction extensions have been introduced to similar effect. For instance, the Integer-FMA extension introduces two instructions that perform FMA-like operations, but on packed integers. Likewise, the 4FMAPS extension allows packing up-to four vectors in one operand (as opposed to a single vector allowed in FMA/IFMA).

Intel processors provide a specialized execution unit for FMA (and its associated sibling) extensions as detailed by the patents~\cite{intel_fma_patent_1,intel_fma_patent_2}. The architecture of the hardware for executing FMA instructions poses an interesting insight: \textit{specialized ``Memory Access Units'' (MAU) reside within the execution cluster of execution engine units} dedicated to FMA instruction execution. Concretely, once the \textit{frontend} unit performs the instruction fetch and instruction decode, it issues the relevant operations to the \textit{backend}, which consists of several units like the register file, the register renaming unit, the scheduler unit, the retirement unit, the execution unit, \textit{and} the memory access unit. We emphasize that both the instruction cache unit as well as the instruction translation look-aside buffer (TLB) unit are a part of the \textit{frontend}, and not the \textit{backend}. Moreover, the MAU interfaces with both the data cache unit as well as the data TLB unit (which sit \textit{outside} the execution engine unit, as a separate memory unit)~\footnote{We omit a detailed architecture description of the FMA execution unit for want of conciseness. Interested readers can refer to Intel patents \cite{intel_fma_patent_1} and \cite{intel_fma_patent_2} for more details. It must also be mentioned that these architectural descriptions depict the overall idea, and the actual implementation in the processors might differ.}. 

\subsection{Leaking arbitrary data from FMA instruction execution}
\label{sec:leaking_fma}

As highlighted in the previous subsection, the FMA execution unit houses MAU whereas the data-cache and dTLB are kept outside the execution unit, in a stand-alone memory unit. We hypothesize that MAUs are essentially meant for \textit{buffering optimizations}. Recall that almost all FMA (and its associated sibling extensions) have instruction variants that can take memory operations. For example, consider \texttt{VFMADD132PD zmm1, zmm2, zmm3}, which multiples the first and third operands, and then adds the product to the second operand. The other flavor of this instruction is \texttt{VFMADD132PD zmm1, zmm2, zmmword PTR [rcx]} which fetches the third operand directly from memory. Since all FMA instructions (and its sibling extensions) have flavors supporting memory operands, we hypothesize the aforementioned ``Memory Access Units'' or MAUs are essentially \textit{buffers} that optimize these memory reads.

From Section~\ref{sec:fp_assists_fma}, we recall that {\sf Shesha} has undiscovered the possibility of speculation in FMA execution. Such speculation, tied with the existence of buffers \textit{inside} FMA execution engine, could leak to leakages. In this context, we test the following two hypotheses related to the transient behavior shown by the FMA instructions:

\begin{description}   \itemsep0em
    \item[\circleB{1}] With speculative execution in FMA execution engine (cf. Section~\ref{sec:fp_assists_fma}), does FMA ``Memory Access Units'' transiently forward data to speculatively executing FMA instructions?
    \item[\circleB{2}] With speculative execution in SSE-AVX execution engine, does FMA ``Memory Access Units'' transiently forward data to speculatively executing SSE-AVX instructions? 
\end{description}

It must be mentioned that hypothesis \circleB{2} draws from our yet another observation about the architecture of FMA execution unit~\cite{intel_fma_patent_1,intel_fma_patent_2}. The register file is shared between SSE-AVX and FMA execution engines. Concretely, the same set of \texttt{xmm/ymm/zmm} registers used by SSE-AVX instructions are also used by FMA instructions. This leads us to investigate whether any \textit{cross-interaction} exists between these two execution engines.

All experiments are performed on $11^{th}$ Gen Intel(R) Core(TM) i5-11500, microcode version \texttt{0x57}, with support for FMA and IFMA extensions. We do not assume any privileges for the adversary, except for the userspace privilege to execute code.

\subsubsection{Intra-FMA Execution Engine Leakage}
\label{sec:intra_fma_leakage}

To test hypothesis \circleB{1}, we execute victim and attacker threads on co-located CPUs. The victim thread executes FMA instructions in a tight loop. We program the victim's memory access to some known byte (specifically \texttt{0xfa}), which is accessed by the FMA instruction. This is done to groom the MAUs within FMA execution engine with known data~\footnote{It must be noted that in an actual attack, the data would be controlled by the victim. We groom the victim's buffer in order to detect the leakage.}.

The attacker process forces speculation in FMA instruction (cf. Section~\ref{sec:fp_assists_fma}) with memory operand. One example is \texttt{VFMADD213PD zmm1, zmm2, zmmword PTR [rcx]}), which multiplies the first and second operands and adds the value loaded by the third operand. Similar to the experiments detailed in Section~\ref{sec:fp_assists_fma}, we force a floating-point assist by grooming \texttt{zmm1} and \texttt{zmm2} with carefully crafted denormal operands. Finally, we encode the entire width of \texttt{zmm1} (which is the destination register) into a Flush+Reload covert channel to leak transiently forwarded data. 

In our experiments, we use the primitives detailed in Table~\ref{tab:reps} to construct and enlarge the transient window to cover speculation in FMA. Overall, we do not observe the programmed victim memory leaking through the adversarial thread's Flush+Reload channel. We attribute this absence of speculative forwarding of stale data to \textcircled{1} lack of speculative forwarding of stale data from MAUs to transiently executing FMA instructions, or \textcircled{2} the inability of constructed transient windows to cover transient execution of FMA instructions (which are inherently complex to execute). There is no evidence to favor one reason over another, though, and we leave this to future work.

\subsubsection{FMA-AVX Execution Engine Leakage}
\label{sec:inter_fma_avx}

To test hypothesis \circleB{2}, we follow a similar setup as Section~\ref{sec:intra_fma_leakage} with the exception that the adversarial thread engages the AVX execution unit. Interestingly, this time, we did observe statistically significant leakage. Listing~\ref{VFMADD213PD_leakage} shows the Flush+Reload channel activity for the victim instruction \texttt{VFMADD213PD zmm1, zmm2, zmmword PTR [rcx]}. Note how the programmed victim memory (i.e. \texttt{0xfa}) is clearly leaking to the adversary.

\begin{lstlisting}[language=C, caption={Leakage when victim executes \texttt{VFMADD213PD zmm1, zmm2, zmmword PTR [rcx]}.}, style=CStyle, label=VFMADD213PD_leakage,  basicstyle=\footnotesize, belowskip= \baselineskip]
<@\textcolor{red}{fa fa fa fa fa fa fa fa fa fa fa fa fa fa fa fa fa fa}@> 7b c8 c6 c8 <@\textcolor{red}{fa fa fa fa}@> 
<@\textcolor{red}{fa fa fa fa fa fa fa fa fa fa fa fa fa fa fa fa fa fa fa fa fa fa fa fa fa fa}@>
<@\textcolor{red}{fa fa fa fa fa fa fa fa fa fa fa fa}@> 22 8c 35 9e ef 76 14 <@\textcolor{red}{fa fa fa fa fa}@>
d4 da bc e4 75 <@\textcolor{red}{fa fa fa fa fa fa fa fa}@> cc 47 5b 11 <@\textcolor{red}{fa fa fa fa}@>
f7 b0 bf <@\textcolor{red}{fa fa fa fa fa fa fa}@> 60 95 ad fc <@\textcolor{red}{fa fa fa fa fa fa fa fa fa fa}@>
<@\textcolor{red}{fa fa fa fa fa fa fa fa fa fa fa fa fa fa}@> 61 6f 40 44 b0 e8 b2 <@\textcolor{red}{fa fa fa}@>
<@\textcolor{red}{fa fa fa fa fa fa fa fa fa fa fa fa fa fa fa fa fa fa fa fa fa fa fa fa fa fa}@>
<@\textcolor{red}{fa fa fa fa fa fa fa fa fa fa fa fa fa fa fa fa fa fa fa fa fa fa fa fa fa fa}@>
\end{lstlisting}

To test the extent of such leakage, we divided the ISEs related to FMA in the following families and tested leakage for all instructions in each family: \textcircled{1} \texttt{VFMADD*} ($12$ instructions from $3$ sub-families: \texttt{VFMADD132*}, \texttt{VFMADD213*}, \texttt{VFMADD231*}), \textcircled{2} \texttt{VFMADDSUB*} ($6$ instructions from $3$ sub-families: \texttt{VFMADDSUB132*}, \texttt{VFMADDSUB213*}, \texttt{VFMADDSUB231*}), \textcircled{3} \texttt{VFMSUB*} ($12$ instructions from $3$ sub-families (\texttt{VFMSUB132*}, \texttt{VFMSUB213*}, \texttt{VFMSUB231*}), \textcircled{4} \texttt{VFMSUBADD*} ($6$ instructions from $3$ sub-families (like \texttt{VFMSUBADD132*}, \texttt{VFMSUBADD213PD*}, \texttt{VFMSUBADD231PD*}), \textcircled{5} \texttt{VFNMADD*} ($12$ instructions from $3$ sub-families like \texttt{VFNMADD132*}, \texttt{VFNMADD213*}, \texttt{VFNMADD231*}), \textcircled{6} \texttt{VFNMSUB*} (contains in-total $12$ instructions from $3$ sub-families like \texttt{VFNMSUB132*}, \texttt{VFNMSUB213*}, \texttt{VFNMSUB231*}), and \textcircled{7} \texttt{IFMA} instructions (\texttt{VPMADD52HUQ} and \texttt{VPMADD52LUQ})\footnote{We present the comprehensive list in Listing~\ref{lst:leaking_instructions} (Appendix~\ref{appx:leaking_instructions}).}.








We have confirmed the leakage in all aforementioned instructions. Moreover, other FMA-related ISEs like \texttt{4FMAPS} also follow a similar architecture as FMA and IFMA and could pose a similar risk of data leakage~\footnote{Due to lack of support for \texttt{4FMAPS} in the CPUs we tested, we could not confirm the existence of leakage in \texttt{4FMAPS} extension family.}.

\subsection{Attack on Cryptographic Applications}
\label{sec:fma_impact}

We now discuss the broader impacts of our findings. While FMA is usually used to accelerate arithmetic, it is the IFMA instruction pair that has received unprecedented adoption in recent years. This is because of IFMA instructions' capability to operate fused multiply and add operations on integers, allowing them to accelerate core operations in traditional cryptography (refer to Intel's guidelines on the same~\cite{fma_guide}). According to Intel Optimization Manual~\cite{intel_opt}, IFMA is introduced for \textit{big number multiplication
for acceleration of RSA vectorized software and other Crypto algorithms (Public key) performance}. However, as a direct consequence of our findings, such optimized implementations are vulnerable to leakages, which we detail next.

\subsubsection{``Almost'' Montgomery Multiplication}
\label{ams52}

In~\cite{drucker2019fast}, the authors suggest using IFMA instructions to speed up modular exponentiation, which forms the basis for several critical cryptographic implementations (include RSA and ECC). This is done in two steps. First, ``almost'' modular multiplication (AMM) is defined algorithmically, in the sense that AMM lacks a final (conditional) reduction step (wrt. tradition modular multiplication). This allows output of one AMM call as inputs to subsequent AMM calls. Next, AMM is implemented and accelerated using \texttt{VPMADD52LUQ} and \texttt{VPMADD52HUQ} instructions. Similar to AMM, the authors in~\cite{drucker2019fast} also propose ``almost'' montgomery squaring (AMS), again accelerated using IFMA instructions. With AMM and AMS in place, modular exponentiation is effectively achieved through \texttt{VPMADD52LUQ}/\texttt{VPMADD52HUQ} instructions.

To test the extent of FMA leakage in AMM/AMS, we implement the \texttt{MulALPart} algorithm from~\cite{drucker2019fast} as depicted in Listing~\ref{lst:mulAPart} (\texttt{\texttt{MulAHPart}} follows suit, but uses \texttt{VPMADD52HUQ} instead of \texttt{VPMADD52LUQ}). Similar to Section~\ref{sec:inter_fma_avx}, the inputs to AMM/AMS are programmed to \texttt{0xfa}. \texttt{MulALPart} performs efficient left-shift in AMM/AMS implementation~\cite{drucker2019fast}, such that this result can directly be used in other IFMA friendly operations in AMM/AMS. Note that AMM/AMS can be performed on any arbitrary integer $A$ exceeding the width of FMA registers, thereby requiring a total of $z$ logical registers (notated as $\{A_{1}, A_{2}, ..., A_{z}\}$). Likewise, $i$ represents the index of the \textit{current} chunk to be operated. $A_{curr}$ is then assigned the broadcasted value of $A[i]$, while $\{X_{1}, X_{2}, ..., X_{i}, ..., X_{z}\}$ are temporary scratch variables. First, as in Line $4$, $X_{i}$ is assigned the value \texttt{VPMADD52LUQ}($X_{i}, A_{curr}, A_{i}$). Then, in a tight loop from $i+1$ to $z$, the temporary products \texttt{VPMADD52LUQ}(\texttt{ZERO}, $A_{curr}$, $A_{i}$) are computed, left-shifted, and added to previously computed $X_{j}$. From our discussion in Section~\ref{sec:leaking_fma}, it is clear that the \textit{third} input operand to IFMA instructions is vulnerable to leakage. With GCC, it is $A_{i}$ that is emitted to be the third operand. This is intuitive: note that in \texttt{VPMADD52LUQ}(\texttt{ZERO}, $A_{curr}$, $A_{i}$), the first two operands are \textit{fixed} for the entire duration of the loop and thus are pre-loaded by the compiler (as first and second operands of \texttt{VPMADD52LUQ}) in registers \texttt{zmm1} and \texttt{zmm3}. Since $A_{i}$ changes in every iteration, this becomes the \textit{third} (memory) operand of \texttt{VPMADD52LUQ}. Therefore, the leaked value is the multiplicand ${A_{i}: i \in \{i+1, i+2, ..., z\}}$, which constitutes the actual input $A$ to AMM/AMS.

\begin{lstlisting}[style=asmstyle, caption={Implementation of MulALPart algorithm~\cite{drucker2019fast}}, label=lst:mulAPart]
.set i, 0
; Initialize <@$X_{i}$@> in zmm0
; Initialize <@$A_{curr}$@> in zmm1
; <@$X_{i}$@> = VPMADD52LUQ <@$X_{i}$@>, <@$A_{curr}$@>, <@$A_{i}$@>
VPMADD52LUQ i(%rcx), %zmm0, %zmm1 
.rept N    ; iteration bounds (i+1, z)
  vpxord %zmm3, %zmm3, %zmm3  ; T = 0
  ; T = VPMADD52LUQ ZERO, <@$A_{curr}$@>, <@$A_{i}$@>
  VPMADD52LUQ i(%rcx), %zmm1, %zmm3
  vpslld $1, %zmm3, %zmm3  ; T = T << 1
  vpaddd %zmm1, %zmm1, %zmm3 ; <@$X_{i}$@> = <@$X_{i}$@> + T
.set i, i+1
.endr
\end{lstlisting}

\begin{lstlisting}[language=C, caption={Leakage when victim executes Listing~\ref{lst:mulAPart}.}, style=CStyle, label=mulAPart_leakage,  basicstyle=\footnotesize, belowskip= \baselineskip]
<@\textcolor{red}{fa fa fa fa fa fa fa fa fa fa fa fa fa fa fa fa fa fa fa fa fa fa fa fa fa}@>
c0 25 a5 a9 1a 26 90 88 <@\textcolor{red}{fa fa fa fa fa fa fa}@> 
30 5d c3 45 fd <@\textcolor{red}{fa fa fa fa fa fa fa fa fa fa fa fa fa fa fa}@>
<@\textcolor{red}{  fa fa fa fa fa fa fa fa fa fa fa}@> 7f f2 6c 11 30 6b <@\textcolor{red}{fa fa fa fa fa fa fa fa}@>
\end{lstlisting}

As depicted in Listing~\ref{mulAPart_leakage}, we successfully observe the programmed input $A = \{A_{1}, A_{2}, ..., A_{z}\}$ leaking in the attacker thread. Implication-wise, this compromises the \textit{input} to AMM/AMS. Depending on the particular use-case where Montgomery multiplication and Montgomery squaring is deployed, such leakage has the potential of compromising sensitive data. For instance, in line with Intel's recommendation~\cite{intel_opt}, should IFMA-enabled Montgomery multiplication/squaring be used to accelerate traditional RSA encryption, then the input to Montgomery multiplication/squaring is the victim plaintext itself. Moreover, Montgomery ladder algorithm is often used as constant-time and side-channel protected RSA and ECC implementations. In such cases, the ability of the attacker to observe the value of the operands of the Montgomery ladder algorithm essentially enables it to distinguish between the if and else part of the control flow~\cite{joye2002montgomery,bernstein2017montgomery}, thereby leaking the secret exponent. Apart from traditional public key algorithms, other forms of cryptosystems also use such multiplication and squaring. We discuss their respective FMA/IFMA accelerations and relevant vulnerabilities next.

\subsubsection{Supersingular Isogeny Diffie-Hellman}
\label{sec:csidh}

In \cite{cheng2021batching}, the authors present IFMA based accelerations of Cumulative Supersingular Isogeny Diffie-Hellman (CSIDH), which is a  post-quantum key establishment scheme in the isogeny-based cryptosystem family. The foundations of this cryptosystem rely on supersingular elliptic curves, which again require core arithemetic operations like multiplications. Intuitively, IFMA seems the perfect choice for accelerating such implementations. The authors propose parallelized operations on the elliptic curve groups, again through optimized implementations of Montgomery multiplication.

In this implementation, we target \texttt{gfp\_mul\_8x1w}\footnote{https://gitlab.uni.lu/APSIA/AVX-CSIDH}, which is a 8-way montgomery multiplication using \texttt{VPMADD52LUQ} and \texttt{VPMADD52HUQ}. As before, we program the inputs to this function to \texttt{0xfa}. Listing~\ref{sidh_leakage} displays the result of the attack.

\begin{lstlisting}[language=C, caption={Leakage when victim executes \texttt{gfp\_mul\_8x1w}.}, style=CStyle, label=sidh_leakage,  basicstyle=\footnotesize, belowskip= \baselineskip]
ac d4 2d 65 <@\textcolor{red}{fa fa fa fa fa fa}@> 6b 22 f7 92 a5 f3 97
<@\textcolor{red}{fa fa}@> 8a fc bf 89 0f <@\textcolor{red}{fa fa fa fa fa fa fa fa}@> ca a4 f9 e8 
a1 60 82 <@\textcolor{red}{fa fa fa fa fa fa fa fa}@> be 78 06 51 61 c2 0a
<@\textcolor{red}{fa fa fa}@> fc <@\textcolor{red}{fa fa}@> 0e <@\textcolor{red}{fa fa fa fa fa fa fa}@> 6a <@\textcolor{red}{fa fa fa fa fa fa fa fa}@>
<@\textcolor{red}{fa fa fa fa fa fa fa fa}@> 9c 2c 32 69 cc bf 9c 2c 32 69 
cc bf 0f 1d 6e b9 04 50 9b 16 83 95 50 fc f6
95 be d0 42 ab <@\textcolor{red}{fa fa fa}@> 05 <@\textcolor{red}{fa}@> bc a7 5b a2 b9 ba 8c d9 
d0 83 10 e1 8d d0 42 ab c4 e4 e3 05 <@\textcolor{red}{fa fa fa}@> 1f e9 
<@\textcolor{red}{fa}@> 9f <@\textcolor{red}{fa}@> c4 <@\textcolor{red}{fa fa fa fa fa fa fa fa fa fa fa fa fa fa fa fa fa fa fa fa fa fa}@>
<@\textcolor{red}{fa fa fa}@> fc <@\textcolor{red}{fa fa}@> 0e <@\textcolor{red}{fa}@> f4 <@\textcolor{red}{fa fa fa}@> c2 <@\textcolor{red}{fa fa fa fa fa fa fa}@>
\end{lstlisting}

\subsubsection{ Supersingular Isogeny Key Encapsulation}

In~\cite{cheng2022highly}, the authors use IFMA to accelerate SIKE, which is a post-quantum isogeny based cryptosystem (similar to SIDH in Section~\ref{sec:csidh}). Here too, IFMA is used to speed upbase/extension field arithmetic, point arithmetic, and
isogeny computations. Since SIKE relies mainly on SIDH-friendly primes, techniques for speeding up modular reduction also apply here. Henceforth, as with Section~\ref{sec:csidh}, this implementation\footnote{https://gitlab.uni.lu/APSIA/AVXSIKE} is also vulnerable to the leakage demonstrated here. 

A slightly different take on accelerating SIKE can be found in~\cite{kostic2019using}, which uses \texttt{VPMADD} (of the FMA family) instead of IFMA, to achieve similar aims of accelerating elliptic curve arithmetic. However, the vulnerability still stands, since \texttt{VPMADD*} sub-family of FMA instructions are also vulnerable to this leak. This fact thereby reinforces the extent of the leakage's impact, irrespective of the underlying cryptosystem and its specification of ISEs used.

\subsubsection{Other Use-Cases}

We briefly mention other use-cases which have similar vulnerability. In~\cite{takahashi2020fast}, the authors propose IFMA based Montgomery multiplication, which again leaks the operands through the leakage described here. Likewise, in~\cite{zheng2023optimized}, Montgomery reduction is also sped up using IFMA, and is vulnerable to our attack. Note that~\cite{zheng2023optimized} speeds up Dilithium, which is a post-quantum digital signature scheme, and has similar vulnerability as AMM in Section~\ref{ams52} of leaking the plaintext message being signed. Finally, recent works like~\cite{boemer2021intel} have also demonstrated the use of IFMA to speed up homomorphic operations as well, and is also vulnerable to the leak\footnote{It is unclear, though, what the extent of exploitability is in the case of homomorphic encryption. This is because even though there is leakage, all data is essentially encrypted and thus secure by the underlying guarantees of the homomorphic encryption scheme.}.

\color{black}

\section{Related Works}    \label{sec:related_works}

Prior works on automated discovery of speculative vulnerabilities in processors can be broadly categorized into two classes based on their approaches for detecting leakages - \textcircled{1} using formal methods and \textcircled{2} fuzzing-based approach. Such tools have their own features and limitations and do not cover the exploratory search space that we target in this work. 

\noindent \textbf{Detection of Leakage through Formal Methods}. Revizor~\cite{oleksenko2022revizor} is a model-based tool built on the fundamentals of \emph{contract traces} in processors and detects violation in the contracts to identify potential leakages.
In order to generate contracts, Revizor uses \emph{random instruction} sequences with use-case specific pruning - straight-line code for meltdown-type violations~\cite{hofmann2023speculation} and conditional branches for spectre-type violations~\cite{oleksenko2023hide}. As the instruction sequences are chosen randomly, both ~\cite{hofmann2023speculation} and \cite{oleksenko2023hide} consider a small subset of x86 instructions such as to not blow-up Revizor's search space. Moreover, Revizor monitors changes in the L1D cache state to detect contract violations. In contrast, \textsf{Shesha} does not rely on random instruction sampling and pruning from apriori knowledge of target speculative behavior. It works upon the formulation of equivalence classes to provide direction to its test case generation which allows defining a ``fitness function'' to maximize occurrences of bad speculation in every equivalence class. Furthermore, \textsf{Shesha} relies on PMCs, instead of L1D cache, to observe if bad speculation occurred. Scam-V~\cite{nemati2020validation, buiras2021validation} is another tool that relies on model-based testing by generating instruction sequences as test-cases to record observation. Recently published Plumber~\cite{ibrahim2022microarchitectural} extends Scam-V to generate leakage templates based on changes in cache state. Both Scam-V and Plumber target ARM-ISA and uses symbolic execution to identify leakages. In contrast, \textsf{Shesha} targets x86-ISA and attempts to find novel leakage paths instead of matching templates.

\noindent \textbf{Detection of Leakage using fuzzing.} Although traditionally, fuzzing has been used to detect bugs in software, hardware-based fuzzing~\cite{domas2017breaking} has seen increasing popularity in recent years. \emph{Transynther}~\cite{moghimi2020medusa} relies upon mutating known code sequences of attacks like~\cite{schwarz2019zombieload,van2019ridl,lipp2020meltdown,canella2019fallout} through fuzzing to discover newer variants of Meltdown-type attacks. 
Similarly, Speechminer~\cite{xiao2019speechminer} generates random code snippets to detect speculation-based attacks. Both these tools use random fuzzing with templates, such as transient window enlargement gadget, faulted load, disclosure gadget, etc. that are known apriori in literature. Since, these tools were developed to target certain classes of speculations, they cannot be directly extended to explore the vast space of bad speculation.
In contrast, \textsf{Shesha} does not place any restrictions on the explored bad speculation classes from apriori decisions. Furthermore, these tools work with limited subset of instructions and do not cover the ISEs handled by \textsf{Shesha}. Similarly, automation tools such as \emph{Osiris}~\cite{weber2021osiris} and \emph{Absynthe}~\cite{gras2020absynthe} explore only contention-based attacks and not transient leakages.

 \color{black}
\section{Mitigations}
\label{sec:mitigations}

We discuss purely software based mitigations against the transient paths discovered by {\sf Shesha}.

\noindent \textbf{Precision based assists and machine clears}: As pointed out in Section~\ref{sec:single_double_precision}, the SMC occuring due to the precision during hardware assist is managed as \textit{loads}/\textit{stores}. Thereby, an \texttt{lfence} before and after an instruction sequence with a homogeneous precision type (either double or single precision) \textit{serializes} the load and eliminates transient execution.

\noindent \textbf{SIMD-Vector Transition based SMC}: As discussed in Section~\ref{sec:simd_vector_transition_smc}, SIMD-Vector transitions based SMC occur to effectively issue \textit{blend} operations. As Intel Optimization Manual suggests~\cite{intel_sdm}, SIMD-Vector based SMC can be prevented by appropriate use of zero-latency instructions \texttt{vzeroupper}/\texttt{vzeroall} before and after a function executing SIMD or Vector instruction sequences.

\noindent \textbf{Assists due to denormal arithmetic}: Intel suggests to set the flags \texttt{FTZ} (Flush to Zero) and \texttt{DAZ} (Denormals are Zero) to prevent denormal numbers from occurring, thereby preventing the floating point assists. However, disabling denormals violates IEEE-754 standard and is not a viable solution for most applications and use-cases~\cite{itsfoss_denormal}. Moreover, custom compiler-level mitigations like~\cite{ragab2021rage} have an unacceptable overhead. 

\noindent \textbf{FMA leakage}: One obvious mitigation is to disable Simultaneous Multi-Threading (SMT), albeit with performance hits. Another (software) mitigation strategy involves using \texttt{lfence} to serialize loads and prevent speculation, transitively preventing the leakage described in this work. An architectural change to disassociate FMA from the SIMD buffer can stop FMA-AVX cross-execution unit leakage.

\section{Conclusion}
\label{sec:conclusion}
Bad Speculation in modern processors causes all in-flight micro-operations to be flushed from the pipeline, contributing to transient execution paths. In this work, we develop an automated transient leakage detection tool, named \textsf{Shesha}, inspired by Particle Swarm Optimization principles. We establish the concept of equivalent classes that represent disjointedly fragmented sub-spaces of bad speculation, which lead to faster convergence and allows us to cover vast search space of ISEs. Using \textsf{Shesha}, we discover novel transient leakage paths and reverse-engineer the root-cause of such transient executions. We discuss how the uncovered speculative execution paths can be used as building blocks for micro-architectural attacks. Finally, we show leakage on cryptographic libraries that use IFMA to accelerate multiplications; we demonstrate how data from FMA execution engine is speculatively forwarded to AVX execution engine. Overall, transient executions are hard to mitigate, and with new optimizations being implemented in next-generation processors, the possibility of such transient paths only seems to be accentuated.

\bibliographystyle{plain}
\bibliography{refs}

\appendix

\section*{Appendix}

\section{{\sf Shesha}: An algorithmic perspective}
\label{sec:algo_description_Shesha}

\begin{algorithm}[!h]
\caption{\small {\sf Shesha driver}}
\label{algo:Shesha2}
\small
\begin{algorithmic}[1]
\Procedure{driver}{}
    \State \texttt{ise\_list} $\leftarrow$ construct\_ise\_list()
    \State \texttt{pos\_vector\_set} $\leftarrow$ construct\_population\_set(\texttt{ise\_list})
    \State cognitive\_phase(\texttt{pos\_vector\_set}, \texttt{ise\_list})
    \State Create sub-swarm set $\mathcal{S}$ from equivalence class of each member
    \State mixed\_phase(\texttt{pos\_vector\_set}, \texttt{ise\_list}, $\mathcal{S}$)
\EndProcedure
\end{algorithmic}%
\end{algorithm}

\begin{algorithm}[!h]
\caption{\small {\sf Shesha}}
\label{algo:Shesha}
\small
\begin{algorithmic}[1]

\Procedure{construct\_ise\_list}{}
    \State \texttt{instructions} $\leftarrow$ parse ISE xml
    \State \texttt{ise\_dict} $\leftarrow \, \{\}$
    \For{\texttt{inst} \textbf{in} \texttt{instructions}}
        \State \texttt{ops} $\leftarrow$ Parse operands for inst
        \State \texttt{ise\_dict}[\texttt{inst}] $\leftarrow$ \texttt{ops}
    \EndFor
    \State \textbf{return} \texttt{ise\_dict}
\EndProcedure

\Procedure{construct\_population\_set}{\texttt{ise\_list}}
    \State \texttt{population\_set} = []
    \For{\texttt{member\_index} \textbf{in} $N$}
        \State \texttt{instruction\_sequence} $\leftarrow$ []
        \For{\texttt{dimension} \textbf{in} $n$}
            \State \texttt{instruction} $\leftarrow$ sample instruction from \texttt{ise\_list}
            \State \texttt{instruction\_sequence}.append(\texttt{instruction})
        \EndFor
        \State \texttt{population\_set}.append(\texttt{instruction\_sequence})
    \EndFor
    \State \textbf{return} \texttt{population\_set}
\EndProcedure

\Procedure{cognitive\_phase}{\texttt{pos\_vector\_set}, \texttt{ise\_list}}
    \State $\alpha$ = 1
    \State $\beta$ = 0.4
    \State $\gamma$ = 0
    \State \texttt{iter} $\leftarrow$ choose iterations
    \While{\texttt{iter} \textbf{is not} 0}
        \For{\texttt{i} \textbf{in} $N$}
            \State \texttt{inst\_prob} $\leftarrow$ [0, 1]
            \If{\texttt{inst\_prob} > $\beta$}
                \State \texttt{d} $\leftarrow$ random sample from $\{1, 2, 3, ...., n\}$
                \State \texttt{pos\_vector\_set}[\texttt{i}][\texttt{d}] $\leftarrow$ sample from \texttt{ise\_list} uniformly at random
            \EndIf
            \State \texttt{operand\_prob} $\leftarrow$ [0, 1]
            \If{\texttt{operand\_prob} > $\beta$}
                \State \texttt{d} $\leftarrow$ random sample from $\{1, 2, 3, ...., n\}$
                \State Mutate operands of \texttt{pos\_vector\_set}[\texttt{i}][\texttt{d}]
            \EndIf
        \State Evalute fitness and assign equivalence class to \texttt{pos\_vector\_set}[\texttt{member\_index}]
        \EndFor
    
    \State \texttt{iter} = \texttt{iter} - $1$
    \EndWhile
\EndProcedure

\Procedure{mixed\_phase}{\texttt{pos\_vector\_set}, \texttt{ise\_list}, $\mathcal{S}$}
    \State $\alpha$ = 1
    \State $\beta$ = 0.1
    \State $\gamma$ = 0.4
    \State \texttt{iter} $\leftarrow$ choose iterations
    \While{\texttt{iter} \textbf{is not} 0}
        \For{\texttt{i} \textbf{in} $N$}
            \State $\mathcal{S}_{i} \, \leftarrow$ determine sub-swarm
            \State \texttt{inst\_prob} $\leftarrow$ [0, 1]
            \State \textbf{Mutate instructions as in cognitive phase}

            \If{\texttt{inst\_swarm\_prob} > $\gamma$}
                \State \texttt{d} $\leftarrow$ random sample from $\{1, 2, 3, ...., n\}$
                \State \texttt{pos\_vector\_set}[\texttt{i}][\texttt{d}] $\leftarrow$ sample from leader of sub-swarm $\mathcal{S}_{i}$
            \EndIf
            
            \State \texttt{operand\_prob} $\leftarrow$ [0, 1]
            \State \textbf{Mutate operands as in cognitive phase}
            
        \State Evalute fitness and assign equivalence class to \texttt{pos\_vector\_set}[\texttt{member\_index}]
        \EndFor
    
    \State \texttt{iter} = \texttt{iter} - $1$
    \EndWhile
\EndProcedure

\end{algorithmic}%
\end{algorithm}

\newpage
\section{SIMD-Vector Transition based SMC}
\label{appx:simd_vector_transient}

\subsection{Non-speculative transition penalty in Micro-architectures prior to Alder Lake and Sapphire Rapids} 

All Intel generation CPUs before Alder Lake and Sapphire Rapids relied upon a combination of \textit{finite state machine} (FSM) based partial register dependency analysis and blend operations to implement the transition. All micro-architectures prior to Alder Lake and Sapphire Rapids define a state of \texttt{ymm} registers termed as \textit{Modified/Unsaved} and implement FSM-based register dependency checks upon transitioning to the use of \texttt{xmm}. For example, consider an instruction \texttt{vpxor ymm1, ymm2, ymm3}, that computes the XOR of \texttt{ymm2} and \texttt{ymm3} and stores the value in $256$-bit register \texttt{ymm1}. Now, suppose a subsequent instruction is \texttt{pxor xmm0, xmm1}, which utilizes the lowermost $128$ bits of the \textit{same} $512$-bit register \texttt{zmm1}, of which the lower $256$ bits (i.e. \texttt{ymm1}) were modified by the \texttt{vpxor} instruction. Then, the register dependency checks (i.e. both \texttt{vpxor} and \texttt{pxor} using the same register) of the FSM shall flag this as a transition from a $256$-bit representation to a $128$-bit representation, wherein the upper half may not be $0$. Hence, blend operations are issued at the issue stage of the pipeline that zero out the upper half of \texttt{ymm1}, before \texttt{pxor} operates upon \texttt{xmm1}~\cite{intel_opt}. We note that FSM operations and issue of blend operations at the \textit{issue} stage of the  pipeline are \textit{not} speculative. This is because FSMs are theoretical models abstracting transitions in system states, and require the inputs to be known in order to implement a state transition. This forces an FSM-based solution to be \textit{non-speculative}, and thus, unable to take advantage of usual speculative optimizations. In fact, because of FSM-based issuing of blend operations, TMA analysis~\cite{intel_opt} involves analyzing the performance counter \texttt{UOPS\_ISSUED.VECTOR\_WIDTH\_MISMATCH} to count the number of such blend operations issued. Moreover, we stress that we do not observe any assists (\texttt{ASSISTS.ANY}) or machine clears (\texttt{MACHINE\_CLEARS.COUNT}), further cementing the lack of transient execution in case FSM decides to issue blend operations.

\subsection{Speculative transition penalty in Alder Lake and Sapphire Rapids}
During our testing campaign using \textsf{Shesha}, we uncover that on Alder Lake and Sapphire Rapids CPUs, SIMD-Vector transitions incur a self-modifying code (SMC) initiated machine clear. This contrasts with zero \texttt{MACHINE\_CLEARS.COUNT} on micro-architectures before Alder Lake and Sapphire Rapids for SIMD-Vector transitions. This implies that among the extensive micro-optimizations done by Intel in transition from Comet Lake to Alder Lake (alternatively from Ice Lake to Sapphire Rapids), one of the optimizations done was to change the transition penalty for SIMD-Vector transitions. We will now investigate this further. From \texttt{MACHINE\_CLEARS.SMC}, it is clear that on Alder Lake and Sapphire Rapids, the FSM based register dependency checks and subsequent issuing of blend operations directly to the pipeline is no longer valid. This is further backed up by the fact that there exists no performance counter event on Alder Lake and Sapphire Rapids to count the number of blend operations (like \texttt{UOPS\_ISSUED.VECTOR\_WIDTH\_MISMATCH} on older generations), implying there is no longer any modifications to the issue state of the pipeline. Additionally, even the TMA analysis for Alder Lake and Sapphire Rapids related to SIMD-Vector transition penalty uses \texttt{ASSISTS.SSE\_AVX\_MIX}, instead of \texttt{UOPS\_ISSUED.VECTOR\_WIDTH\_MISMATCH} as with previous generations like Comet Lake (as depicted by \texttt{pmu-events} in Linux source code~\cite{linux_kernel}).

Interestingly, we uncover that the blend operations are now inserted directly into the instruction cache on Alder Lake and Sapphire Rapids because of the self-modifying code machine clear. To verify this, we rely upon metrics for instruction cache (precisely \texttt{ICACHE.ACCESSES} and \texttt{ICACHE.MISSES}, which count the total number of requests and total number of misses wrt. the instruction cache respectively). Averaged across $1$ million iterations, we observe \texttt{ICACHE.ACCESSES} for a code sequence with no SIMD-Vector SMC machine clear to be $21$ while it climbs to $70$ for code sequences with SIMD-Vector SMC machine clear, implying (on average) about $49$ more instruction cache accesses occurred. However, \texttt{ICACHE.MISSES} were $0$ in both cases, implying that even though around $49$ additional instruction cache accesses are needed for the SIMD-Vector SMC machine clear, \textit{none} of them are cache misses. This observation corroborates with the known general understanding of SMC, i.e.,  any self-modifying code \textit{writes} to the code section of the binary, and is implemented as \textit{store} operations on the instruction cache. Since we observe $0$ \texttt{ICACHE.MISSES}, it is not far-fetched to speculate that the blend operations due to SIMD-Vector transition are directly written to the instruction cache, which in turn causes bad speculation. The pipeline needs to be re-steered according to the newly written instructions in the instruction cache. This becomes the root cause of the novel transient execution path uncovered by {\sf Shesha}.

Furthermore, our analysis shows that the SMC machine clear occurs irrespective of \textcircled{1} whether the \texttt{ymm} registers are in \textit{Modified/Unsaved} state (i.e. the upper $128$ bits of the $256$-bit \texttt{ymm} register) are clean, as well as \textcircled{2} whether there is actually any register dependency upon SIMD-Vector transitions. This cements the fact that the SIMD-Vector transition SMC is indeed a \textit{pessimistic transition penalty-} penalty incurred irrespective of whether there is register dependency or whether \texttt{ymm} register is indeed unclean. Interestingly, incurring transition penalty without actually checking for these conditions is also done by AMD in its recent EPYC processors, where spilling occurs on the higher $128$ bits of \textit{all} \texttt{ymm} registers, irrespective of which \texttt{ymm} register is involved in the transition (refer Section~$2.11.6$ in \cite{amd_epyc}). 

\section{Precision based Transient Execution Paths}
\label{appx:precision_transient}

Let us represent the sub-classes of single precision and double precision floating point instructions as \textbf{SP} and \textbf{DP}.

\subsection{Precision based Hardware Assist}
\label{sec:precision_hardware_assist}

A hardware assist (represented by \texttt{ASSISTS.HARDWARE} in Table~\ref{tab:pmc}) is required when an exceptional scenario arises where dedicated hardware extensions are needed to execute the concerned instruction. For performance optimizations, Intel provides dedicated hardware for a variety of operations. To exemplify a few, there exist \texttt{AES-NI} execution units for cryptographic operations, hardware support for half-precision conversion operations using \texttt{VCVTPS2PH}, and so on. While the programmer can interface with these hardware extensions separately through ISE instructions, a dedicated hardware assist is needed when an exceptional situation arises where intervention of such hardware is required. One such scenario unearthed by \textsf{Shesha} is the requirement of hardware assists in converting representations of floating point numbers. Without loss of generality, consider \texttt{DP xmm0, xmm1, xmm2} followed by \texttt{SP xmm3, xmm4, xmm0}. This instruction sequence updates \texttt{xmm0} to contain a double-precision floating-point number.  However, without any explicit conversion instruction in between, the subsequent instruction \texttt{SP xmm3, xmm4, xmm0} is forced to rely upon a hardware assist to \textit{convert} \texttt{xmm0}'s natural double-precision floating point representation to a single-point representation, thereby creating a transient execution window.

\subsection{Precision based SMC Machine Clear}
\label{sec:smc_machine_clear}

As evident from Section~\ref{sec:precision_hardware_assist}, mismatches in the precision of a register involved in a RAW data dependency incur hardware assists for fixing the mismatching representations. However, considering \texttt{DP xmm0, xmm1, xmm2} followed by \texttt{SP xmm3, xmm4, xmm0}, we note that without the presence of an explicit conversion operation (like \texttt{CVTPD2PS} or \texttt{CVTPS2PD}) between the double-precision instruction \textbf{DP} and the single-precision instruction \textbf{SP}, the processor is faced with two challenges. First, it needs a way to represent \texttt{xmm0} as single-precision floating-point value consumable by instruction \textbf{SP} (already detailed in Section~\ref{sec:precision_hardware_assist}). Secondly, for architectural correctness, it needs a way to \textit{retain} the initial value of \texttt{xmm0} in double-precision at the end of execution of \textbf{DP}, even though a corresponding single-precision representation is needed for \textbf{SP}.

Note that architecturally, instruction sequence \texttt{DP xmm0, xmm1, xmm2} followed by \texttt{SP xmm3, xmm4, xmm0} \textit{requires} \texttt{xmm0} to be still double precision \textit{after} \textbf{SP} ends execution. Hence, architecturally, this code sequence must behave as if two implicit conversion instructions were written into the code sequence by the programmer- \texttt{DP xmm0, xmm1, xmm2}, followed by an explicit single precision conversion \texttt{CVTPD2PS xmm0, xmm0}, followed by the single precision \texttt{SP xmm3, xmm4, xmm0}, and then followed by an explicit conversion back to double precision \texttt{CVTPS2PD xmm0, xmm0}. Since the programmer has not included these explicit conversions, the hardware assist described in Section~\ref{sec:precision_hardware_assist} cannot guarantee architectural correctness \textit{without} auxiliary code to manage \texttt{xmm0}'s initial double precision value, and restoring it later. This management code is exactly what the SMC machine clear is performing in this context. To ensure this further, we rely upon instruction cache metrics \texttt{ICACHE.ACCESSES} and \texttt{ICACHE.MISSES}, counting the number of total accesses to the instruction cache and corresponding misses respectively. Averaged across 1 million executions, we observe \texttt{ICACHE.ACCESSES} to be $21$ for a \textbf{SP}-\textbf{SP} or a \textbf{DP}-\textbf{DP} instruction sequence, while it climbs to $36$ for \textbf{SP}-\textbf{DP} or \textbf{DP}-\textbf{SP} instruction sequences. Moreover, we observe \textit{no} \texttt{ICACHE.MISSES} even though (on average) $15$ more instructions are added to the instruction cache as a result of the SMC. This reinforces the idea that all instructions needed to manage \texttt{xmm0}'s double precision state while executing a single precision operation are directly added to the instruction cache. And these newly added instructions force re-steering the pipeline, squashing any in-flight instructions, and re-issuing these instructions. This becomes the root cause of the precision-based SMC machine clear, thereby creating yet another novel transient execution path.

\subsection{Precision based Memory Ordering}
\label{sec:precision_memory_ordering}

In Section~\ref{sec:precision_hardware_assist} and Section~\ref{sec:smc_machine_clear}, we detailed how two separate and novel transient executions paths are uncovered by \textsf{Shesha} in order to manage the state of a SIMD register upon \textit{implicit} precision conversions. We note that simply from observing an SMC machine clear (cf. Section~\ref{sec:smc_machine_clear}), it is not possible to uncover what exact instructions are being added to the instruction cache. However, yet another transient execution path uncovered by \textsf{Shesha} offers more insights into the exact nature of this SMC - occurrence of \textit{memory ordering} issues during \textbf{SP}-\textbf{DP} or \textbf{DP}-\textbf{SP} without any explicit conversions. We delve deeper into such memory ordering in this section.

Again, without loss of generality, we consider the instruction sequence \texttt{DP xmm0, xmm1, xmm2} followed by \texttt{SP xmm3, xmm4, xmm0} incurring a memory ordering machine clear. Concretely, since both \textbf{SP} and \textbf{DP} are exclusively \textit{arithmetic} instructions, the only possibility of a memory ordering issue arises if the SMC is relying upon \textit{load}/\textit{store} operations to manage the initial and final state of \texttt{xmm0} (cf. Section~\ref{sec:smc_machine_clear}). This hypothesis is further concretized by the fact that memory ordering issues in Intel CPU occur due to two reasons as stated in~\cite{intel_perfmon}- \textcircled{1} a \textit{load} instruction does not conform to \texttt{x86} memory ordering rules or \textcircled{2} initiation of MESI snoop cache coherence protocol. In this case, our observation and experiment design indicates \textit{no} MESI initiations occur, thereby leaving the first case as the root-cause of memory ordering issue. To elaborate further, we observe zero snoop requests through measuring performance counters like \texttt{MEM\_LOAD\_L3\_HIT\_RETIRED.XSNP\_HIT}, \texttt{MEM\_LOAD\_L3\_HIT\_RETIRED.XSNP\_NO\_FWD}, \texttt{OCR.DEMAND\_DATA\_RD.L3\_HIT.SNOOP\_HITM}, and other counters involving snoop requests~\cite{intel_perfmon}, leaving only the possibility of a faulty \textit{load} as the potential cause of the memory ordering issue. Moreover, should we insert an explicit \texttt{lfence} to obtain the sequence: \texttt{DP xmm0, xmm1, xmm2}, followed by \texttt{lfence} and then followed by \texttt{SP xmm3, xmm4, xmm0}, the memory ordering issues \textit{vanishes}. Combining both these observations (\textcircled{1} absence of MESI snoop requests and \textcircled{2} serialized \textit{load} executions using \texttt{lfence} preventing memory ordering issues), we conclusively state that the underlying SMC to manage internal state of \texttt{xmm0} utilizes at least one \textit{store}/\textit{load} pair, where the \textit{load} is speculatively issued before the \textit{store} causing a memory ordering issue. We stress that the offending \textit{store}/\textit{load} pair is on the \textit{same} effective virtual address because we observe no memory disambiguation issues (through \texttt{MACHINE\_CLEARS.DISAMBIGUATION} in Table~\ref{tab:pmc}). This is because memory disambiguation issues arise due to a \textit{store}/\textit{load} pair on actual differing effective virtual addresses being speculated to operate upon the \textit{same} effective virtual address during the address resolution process.

Now that we have established occurrence of \textit{load}/\textit{store} operations during implicit precision conversions, we investigate subsequent behaviour of these operations. To do so, we rely on the following performance counters: \texttt{UOPS\_ISSUED.ANY}, \texttt{IDQ.MS\_UOPS}, and \texttt{UOPS\_RETIRED.MS}. Here, \texttt{UOPS\_ISSUED.ANY} counts the total number of micro-operations issued by the Resource Allocation Table (RAT) to the Reservation Stations (RS) and provides a metric to understand the \textit{overall} micro-operations in a code sequence. Likewise, \texttt{IDQ.MS\_UOPS} counts the number of micro-operations serviced from the microcode sequencer and is helpful to compute the fraction of \texttt{UOPS\_ISSUED.ANY} micro-operations coming from the microcode ROM. Similar to \texttt{IDQ.MS\_UOPS}, \texttt{UOPS\_RETIRED.MS} counts the number of micro-operations (out of \texttt{IDQ.MS\_UOPS}) that are actually retired, and gives an understanding of any micro-operation serviced by the microcode sequencer that was not architecturally committed. Averaged across 1 million iterations, we observe \texttt{UOPS\_ISSUED.ANY} to be $80$ (with no memory ordering machine clear) as against $125$ (with memory ordering). Likewise, both \texttt{IDQ.MS\_UOPS} and \texttt{UOPS\_RETIRED.MS} are on average $45$ and $43$ in the absence of memory ordering, but climb to $76$ and $54$ in the presence of memory ordering, respectively. These numbers, as expected, point to an increased involvement of the microcode sequencer when memory ordering issues occur. However, we do not observe any perceivable difference in activity in the L1 data cache using relevant counters from~\cite{intel_perfmon}.

Absence of any L1 data cache activity in spite of the increased activity of the microcode sequence strongly suggests upon the involvement of an on-core buffer to temporarily hold the results of \textit{store} operations, to be later fetched by \textit{load} operations. Indeed, recent works like~\cite{moghimi2023downfall} have uncovered previously undisclosed SIMD temporal buffers in Intel micro-architectures, which are utilized by SIMD/Vector \textit{store}/\textit{load}/\textit{gather}/\textit{scatter} operations. To further investigate this, we perform the \textit{Store Bound} category TMA analysis of memory bound micro-operations~\cite{intel_opt}. According to Intel Software Development Manual~\cite{intel_opt}, \textit{store operations
are buffered and executed post-retirement due to memory ordering requirements}. The extent of this buffering can thereby be observed by analyzing the number of clock cycles with \textit{low} execution port utilization, thereby hinting how many clock cycles were spent in non-executable tasks (in our case, buffering of \textit{store} operations). To do so, we rely upon the performance counter \texttt{EXE\_ACTIVITY.EXE\_BOUND\_0\_PORTS}, which measures the number of clock cycles wherein \textit{no} micro-operations were executed on \textit{all} execution ports, even though \textcircled{1} at least one reservation station was full (implying there was a micro-operation waiting to be executed), \textcircled{2} Store Buffer was not full (implying there was no outstanding blocking \textit{store}), and \textcircled{3} there was no outstanding load. In other words, \texttt{EXE\_ACTIVITY.EXE\_BOUND\_0\_PORTS} measures cycles that were spent on other tasks (like buffering \textit{stores}) even though valid micro-operations were waiting to be executed. In our context, averaged across 1 million iterations, we observe the average number of clock cycles reported by \texttt{EXE\_ACTIVITY.EXE\_BOUND\_0\_PORTS} to be $9$ (without memory ordering machine clear), as opposed to $16$ (with memory ordering machine clear). This clearly indicates an increase in the \textit{Store Bound} tasks, and to the highly probable use of a SIMD buffering of \textit{stores}. However, we cannot conclusively state whether store buffering occurring in precision-based memory ordering issues relies on the same SIMD buffer uncovered by~\cite{moghimi2023downfall} or some other temporal buffer.

\section{SMT Covert Channel Sender/Receiver}
\label{sec:smt_covert_channel_code}

\begin{algorithm}[!h]
\caption{\small SMT Covert Channel sender}
\label{algo:covert_channel_sender}
\small
\begin{algorithmic}[1]

\Procedure{sender}{$\mathcal{C}$}
    \State Pin to logical core $i$ on physical core $k$
    \State Sample bit \texttt{b} $\leftarrow \{0, 1\}$
    \If{\texttt{b} = 1}
            \State Trigger transient execution (eg. SIMD-Vector Transition, Precision intermixing, FMA assist, AES-SIMD assist etc)
            \State Trigger execution of code sequence $\mathcal{C}$
    \Else
            \State Trigger execution of code sequence $\mathcal{C}$
    \EndIf
\EndProcedure

\end{algorithmic}%
\end{algorithm}

\begin{algorithm}[!h]
\caption{\small SMT Covert Channel receiver}
\label{algo:covert_channel_receiver}
\small
\begin{algorithmic}[1]

\Procedure{receiver}{$\mathcal{C}$}
    \State Pin to logical core $j$ on physical core $k$
    \State \texttt{start\_time} $\leftarrow$ \texttt{rdtscp()}
    \State Trigger execution of code sequence $\mathcal{C}$
    \State \texttt{end\_time} $\leftarrow$ \texttt{rdtscp()}
    \If{\texttt{end\_time - start\_time} > \texttt{threshold}}
            \State \texttt{received\_bit} = $1$
    \Else
            \State \texttt{received\_bit} = $0$
    \EndIf
\EndProcedure

\end{algorithmic}%
\end{algorithm}

\section{List of leaking Instructions}
\label{appx:leaking_instructions}

We capture a comprehensive list of instructions for which we could establish the leakage.

\begin{lstlisting}[style=asmstyle, caption={FMA/IFMA instructions that leak.}, label=lst:leaking_instructions]
      ; FMA instructions
VFMADD132PD VFMADD132PS VFMADD132SD VFMADD132SS 
VFMADD213PD VFMADD213PS VFMADD213SD VFMADD213SS 
VFMADD231PD VFMADD231PS VFMADD231SD VFMADD231SS 
VFMADDSUB132PD  VFMADDSUB132PS VFMADDSUB213PD 
VFMADDSUB213PS VFMADDSUB231PD  VFMADDSUB231PS 
VFMSUB132PD VFMSUB132PS VFMSUB132SD VFMSUB132SS    
VFMSUB213PD VFMSUB213PS VFMSUB213SD VFMSUB213SS    
VFMSUB231PD VFMSUB231PS VFMSUB231SD VFMSUB231SS    
VFMSUBADD132PD  VFMSUBADD132PS VFMSUBADD213PD 
VFMSUBADD213PS  VFMSUBADD231PD VFMSUBADD231PS 
VFNMADD132PD VFNMADD132PS VFNMADD132SD 
VFNMADD132SS VFNMADD213PD VFNMADD213PS 
VFNMADD213SD VFNMADD213SS VFNMADD231PD 
VFNMADD231PS VFNMADD231SD VFNMADD231SS   
VFNMSUB132PD VFNMSUB132PS VFNMSUB132SD 
VFNMSUB132SS VFNMSUB213PD VFNMSUB213PS 
VFNMSUB213SD VFNMSUB213SS VFNMSUB231PD 
VFNMSUB231PS VFNMSUB231SD VFNMSUB231SS   
     
     ; IFMA instructions
VPMADD52HUQ     VPMADD52LUQ
\end{lstlisting}

\end{document}